\documentclass[twocolumn,showpacs,preprintnumbers,amsmath,amssymb,pre]{revtex4-1}

\usepackage{graphicx}
\usepackage{dcolumn}
\usepackage{bm}
\usepackage{xcolor}

\begin{document}

\title{Suppression of spatially periodic patterns by dc voltage}

\author{N\'{a}ndor \'{E}ber$^1$,
P\'{e}ter Salamon$^1$,  Bal\'{a}zs Andr\'{a}s Fekete$^1$, Ridvan Karapinar$^2$, Alexei Krekhov$^3$, and
\'{A}gnes Buka$^1$ }

\address{$^1$ Institute for Solid State Physics and Optics,\\
Wigner Research Centre for Physics, Hungarian Academy of
Sciences,\\ H-1525 Budapest, P.O.B.49, Hungary}

\address{$^2$ 100. Yil University, Department of Physics, 650580 Van,
Turkey}

\address{$^3$ Max Planck Institute for Dynamics and
Self-Organization, 37077 G\"{o}ttingen, Germany}


\date{\today}

\begin{abstract}
The effect of superposed dc and ac applied voltages on two types of spatially periodic instabilities in nematic liquid crystals, flexoelectric domains (FD) and electroconvection (EC), was studied.
The onset characteristics, threshold voltages and critical wave vectors, were determined.
We found that in general the superposition of driving with different time symmetries inhibits the pattern forming mechanisms for FD and EC as well.
As a consequence the onset extends to much higher voltages than the individual dc or ac thresholds.
A dc bias induced reduction of the crossover frequency from the conductive to the dielectric EC regimes and a peculiar transition between two types of flexodomains with different wavelengths were detected.
Direct measurements of the change of the electrical conductivity and its anisotropy, induced by the applied dc voltage component, showed that the dc bias substantially affects both parameters. Taking into account the experimentally detected variations of the conductivity in the linear stability analysis of the underlying nemato-hydrodynamic equations, a qualitative agreement with the experimental findings on the onset behaviour of spatially periodic instabilities was obtained.

\end{abstract}



\maketitle

\section{\label{sec:intro} \bf{Introduction}}

Nematic liquid crystals (nematics) are characterized by a long range orientational ordering described by the director field $\mathbf{n}$, thus representing anisotropic fluids.
The preferred orientation $\mathbf{n}$ can easily be changed by an electric field $\mathbf{E}$ \cite{Blinov,deGennes}. This process is mostly
governed by the typically dominant quadratic dielectric
contribution $f_{\mathrm{d}}=-\frac{1}{2}
\varepsilon_{0} \varepsilon_{\mathrm{a}} (\mathbf{n E})^2$ to the free
energy of the system. Here $\varepsilon_{0}$ is the electric
constant and $\varepsilon_{\mathrm{a}} = \varepsilon_{\|}-
\varepsilon_{\bot}$ is the dielectric anisotropy (the difference
of permittivities measured along and perpendicular to $\mathbf{n}$).
The linear with respect to $\mathbf{E}$ flexoelectric contribution of $f_{\mathrm{fl}}=-(\mathbf{P}_{\mathrm{fl}} \mathbf{E})$ may also be
important \cite{Eberbook}. Here $\mathbf{P}_{\mathrm{fl}} = e_1 \mathbf{n} (\mathrm{div}
\mathbf{n}) - e_3 \mathbf{n} \times (\mathrm{curl}
\mathbf{n})$ is the flexoelectric polarization induced by splay and
bend director deformations, with $e_1$ and $e_3$ being the relevant
flexoelectric coefficients.

Though a nematic is electrically neutral in its basic state, an inhomogeneous electric charge density $\varrho_e$ may develop in the electric field due to the director deformations; then a Coulomb force $\mathbf{F}=\varrho_{\mathrm{e}}
\mathbf{E}$ arises, which may induce a material flow
$\mathbf{v}$.

In most experiments and applications thin (5--20 $\mu$m) nematic layers are sandwiched between transparent electrodes covered by aligning layers, which ensure a uniform quiescent state.
Upon applying an electric voltage $U$ to the electrodes, director
distortions may occur if $U$ exceeds some critical value $U_{\mathrm{c}}$
\cite{Blinov}. In many cases, the distortions are uniform in the
plane parallel to the substrates. These electro-optical effects
are utilized in liquid crystal displays (LCDs).

Under some conditions the applied voltage
may induce patterns. Two basic types of patterns can be
distinguished: equilibrium director deformations, spatially periodic in the plane of the nematic layers with a wave vector
$\mathbf{q}$, and dissipative ones, where director distortions are
accompanied by vortex flow and are therefore called
electroconvection (EC). In this latter phenomenon electrical
conductivity plays a crucial role.

Flexoelectric domains (shortly flexodomains, FD) induced by dc voltage are a paradigm
of the first type. They have been observed by polarizing
microscope in planar nematics (initial homogeneous director orientation $\mathbf{n}_0$ is parallel to the confining plates) as a series of dark and bright
stripes running parallel to $\mathbf{n}_0$
($\mathbf{q}_{\mathrm{FD}} \, \bot \, \mathbf{n}_0$) \cite{Vistin1970}. The
pattern has been interpreted as a flexoelectricity induced
periodic director modulation \cite{Bobylev1977}: it appears when
the free energy reduction due to the flexoelectric polarization
overcomes the elastic and dielectric increase of the free energy.
This may only occur if certain relations between the material
parameters of the nematic fulfils \cite{Krekhov2011}.

Electroconvection (EC) being of dissipative origin may also yield stripe (roll) patterns
\cite{Kramer1996}, however, the orientation of the rolls is
different from that of FD: the rolls are either perpendicular to $\mathbf{n}_0$ (normal rolls) or
run at an angle (oblique rolls). The obliqueness is usually characterized by the angle $\alpha$ between $\mathbf{q}$ and $\mathbf{n}_0$ (typically $0^{\circ} \leq \alpha \lesssim 45^{\circ}$). The most common example of
electroconvection (standard EC) is observed in planar nematics
having a negative dielectric anisotropy $\varepsilon_{\mathrm{a}} <0$ and a positive electrical conductivity
anisotropy $\sigma_{\mathrm{a}} = \sigma_{\|} - \sigma_{\bot} >0$, where $\sigma_{\|}$ and $\sigma_{\bot}$ are the electrical conductivities measured along and perpendicular to $\bf n$. The
driving feedback mechanism, a coupling of spatial director fluctuations to space charge separation and circulation flow, was invented by Carr \cite{Carr1969} and Helfrich \cite{Helfrich1969}.
For the applied voltage $U$ below some critical value $U_{\mathrm{c}}$ the fluctuations decay, however, they grow to a
finite amplitude pattern for $U>U_{\mathrm{c}}$. The periodic director modulation acts as an optical grating, such that the pattern is easily visualized.

The effects mentioned above can be induced by dc
($U_{\mathrm{dc}}$) as well as by ac ($U_{\mathrm{ac}}$) applied voltages. In the
latter case $U_{\mathrm{ac}}$ corresponds to the rms value of the driving
sinusoidal voltage $U = \sqrt{2} U_{\mathrm{ac}} \sin(\omega t)$ of
frequency $f$; here $\omega=2 \pi f$ is the circular frequency. It
was found that the critical (threshold) voltages are frequency
dependent and the shape of the $U_{\mathrm{c}}(f)$ curves as well as the
$f$-range of existence depend strongly on the type of the pattern
\cite{Kramer1996}. For example, for a typical set of material
parameters, FD are seen only at low $f$ below a few Hz, while EC
patterns might be observed up to several ten kHz.

By now theoretical description has been
developed to describe the patterns discussed
above, which is known as the standard model of electroconvection
including flexoelectricity \cite{Krekhov2008,Krekhov2011} (we will
refer to it as the extended SM).
The linear stability analysis of the extended SM provides the threshold $U_{\mathrm{c}}(f)$, the
critical wave vector $\mathbf{q}_{\mathrm{c}}(f)$ and the spatio-temporal
dependence of the material flow $\mathbf{v}(\mathbf{r},t)$, the electric potential, and the director $\mathbf{n}(\mathbf{r},t)$ at the
onset, in a good agreement with experiments \cite{TothKatona2008}. The
equations have solutions of three different types. One is at dc driving (we will refer to it as the
\emph{dc-mode}): in this case $\mathbf{n}(\mathbf{r})$,
$\mathbf{v}(\mathbf{r})$ and $\varrho_{\mathrm{e}}(\mathbf{r})$ are time
independent. The other two occur at ac driving.
For frequencies $f$ below the crossover frequency $f_{\mathrm{c}}$ (in the {\em conductive mode}), in leading order, the director and the velocity are stationary, while the charge density oscillates with $f$.
For $f > f_{\mathrm{c}}$ (in the {\em dielectric mode}), in contrast, $\mathbf{n}$ and $\mathbf{v}$ oscillate with the ac frequency in leading order, while $\varrho_e$ is stationary. It should be noted that due to these differences in
time symmetries there is no smooth transition from the ac $f
\rightarrow 0$ limit to the dc case \cite{Krekhov2011,Eber2012}.

The complexity of pattern types and their temporal behaviour
justify a special attention to pattern formation
driven by a superposition of two voltages ($U_1$ + $U_2$), which
in themselves would induce patterns of different types. At such a combined driving, the key questions are:
where is the limit of stability of the initial homogeneous state
in the $U_1$--$U_2$ plane and how does the pattern morphology
change when moving along this stability limit curve (SLC).

In pioneering works \cite{John2004,Heuer2006,Pietschmann2010}, EC under the superposition of a
high ($f_1$) and a low ($f_2$) frequency ac voltages with an integer frequency ratio  ($f_1 : f_2=4:1 - 2:1$) was
studied and at some voltage
combinations formation of subharmonic patterns was detected. The
effect of mixing two ac voltages with an arbitrary frequency
ratio, however, has not been studied yet.

Another interesting case is the superposition of ac and dc
voltages, which by symmetry reasons, as we have pointed out above,
is not equivalent with the superposition of two ac voltages of
frequencies $f_1$ and $f_2\rightarrow 0$.
Experimental studies on electro-optics in nematics at combined ac+dc driving
are so far very scarce. E.g., the dc threshold of
FD was found to increase upon superposing ac voltage in a
calamitic \cite{Marinov2006} as well as in a bent-core nematic
\cite{Tadapatri2012}; however, the instabilities at pure ac driving or at a small dc bias voltage have not been studied. Combined driving was also applied for the
nematic 5CB (4-pentyl-4'-cyanobiphenyl) having $\varepsilon_{\mathrm{a}}
>0$ and $\sigma_{\mathrm{a}} >0$ and thus exhibiting nonstandard EC \cite{Kumar2010}. The
superposition of ac and dc voltages resulted in changes of the
thresholds and also in appearance of new pattern morphologies
\cite{Aguirre2012}. The influence of the combined driving on secondary EC instabilities has been reported in \cite{Batyrshin_2012} where the effect of spatiotemporal synchronization of oscillating zig-zag EC rolls under an increase of a dc bias voltage was found.

From theoretical point of view, the consequences of the combined
ac+dc driving are non-trivial, since the underlying nemato-hydrodynamic equations contain terms linear
as well as quadratic in the applied voltage. Thus even in the linear stability analysis, the solution may not be obtained as a
simple superposition of the three basic modes of different time
dynamics. Recently, using the extended SM, the SLC as well as
the wave vector $\mathbf{q}$ along the SLC have been
calculated for four basic cases when changing from pure dc to pure ac voltage \cite{Krekhov2014}: (A) dc-mode of EC at dc voltage and conductive EC mode at ac voltage with $f < f_{\mathrm{c}}$; (B) dc-mode of EC at dc voltage and dielectric EC mode at ac voltage with $f > f_{\mathrm{c}}$; (C) FD at dc voltage and conductive EC mode at ac voltage with $f < f_{\mathrm{c}}$; (D) FD at dc voltage and dielectric EC mode at ac voltage with $f > f_{\mathrm{c}}$.
The calculations predicted closed SLC connecting the
ac and dc threshold voltages for all cases. In case A the obtained
SLC was smooth and convex, the superposed threshold was always
lower than the pure ac or dc ones, and $\mathbf{q}$ varied
continuously from its ac to the dc value. In contrast to that, in
cases B to D the SLC exhibited a protrusion toward larger dc
voltages.
Moreover, in those cases the SLC had a break indicating a sharp
transition between patterns with different magnitude and/or
direction of the wave vectors.

The above theoretical results have been compared with experiments, which explored the influence of combined ac+dc driving on standard EC in a nematic mixture (Merck Phase 5) \cite{Krekhov2014,Salamon2014}. A comparison of experimental data
with theoretical predictions yielded a good match for case A, however,
only at the lowest frequencies \cite{Krekhov2014}. At higher
frequencies or in cases B to D, measurements indicated a
substantial increase of the ac threshold upon superposing dc
voltage;
an effect becoming more pronounced with increasing the frequency
of the ac component. In extreme cases (in the dielectric regime)
it resulted in the 'opening' of the SLC: the superposed ac and dc
voltage reached the upper limit of the voltage source without
inducing pattern, even though the voltage(s) exceeded several
times the pure ac or dc thresholds (see Fig.~1 of \cite{Salamon2014}).
Later, in Section \ref{sec:discussion}, we provide an explanation of
this pattern inhibition.

Our measurements were
performed using a nematic liquid crystals exhibiting standard EC at ac
driving and FD at dc voltage, thus allowing an experimental check of the theoretical
predictions for cases C and D.

The paper is organized as follows. After introducing our
set-up, experimental methods, and the studied compound in
Section~\ref{sec:setup}, we present our results (experimental as well
as theoretical) in Section~\ref{sec:results}, reporting on various
pattern forming scenarios and on electric current measurements.
These  results are further analyzed in Section~\ref{sec:discussion}
and the paper is finally closed with conclusions in
Section~\ref{sec:sum}.

\section{\label{sec:setup} \bf{Compounds, experimental set-up
and evaluation method} }

\begin{figure}[!h]
\begin{center}
\includegraphics[width=8cm]{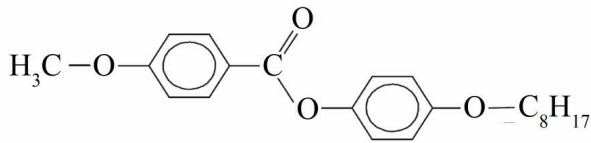}
\end{center}
\caption{The chemical structure of the nematic 4-n-octyloxyphenyl
4-n-methyloxybenzoate (1OO8).} \label{fig:1oo8}
\end{figure}

The compound selected for the studies, 4-n-octyloxyphenyl
4-n-methyloxybenzoate (1OO8), has nematic phase in the
temperature range between 53 $^{\circ}$C and 77 $^{\circ}$C
\cite{Salamon2013}. Its chemical structure is shown in Fig.~\ref{fig:1oo8}.
It has negative dielectric and positive conductivity anisotropies;
hence standard EC develops at pure ac driving. At pure dc voltage, depending on the sample thickness $d$ and on the electrical conductivity, it exhibits either flexodomains \cite{Salamon2013} or standard EC.

The liquid crystal was filled into commercial cells ($d=19.5$
$\mu$m, WAT, Poland  and $d=50$ $\mu$m, EHC. Co, Japan); rubbed
polyimide coatings on the transparent electrodes ensured a planar
initial alignment of the nematic. The cell was placed
into a temperature controlled compartment (a Linkam
LTS350 hot-stage with a TMS 94 controller); experiments were
performed at the temperature $T=58\pm0.05$ $^{\circ}$C, i.e., well in
the nematic phase range.

The voltage applied to the cell was provided by the function
generator output of a TiePie Handyscope HS3 digital oscilloscope
through a high voltage amplifier. It allowed to synthesize the
driving voltage in the form of $U = U_{\mathrm{dc}} + \sqrt{2} U_{\mathrm{ac}}
\sin(\omega t)$ with arbitrary $U_{\mathrm{dc}}/U_{\mathrm{ac}}$ ratios.
The two oscilloscope channels of the same device were used to
record simultaneously the signals of the applied voltage and the
current flowing through the cell, thus providing information on
the sample impedance during pattern formation.

The voltage-induced patterns were observed by a polarizing
microscope (Leica DM RXP) at white light
illumination using a single polarizer (shadowgraph
\cite{Rasenat1989} technique). An attached digital camera,
Mikrotron EoSens MC1362, was used to record
snapshot sequences for documentation and/or later digital
processing. It was capable of high speed imaging at a variable (maximum
2000 frames/s) rate with a spatial resolution of 520*512 pixels at
256 grey levels. A specially designed trigger logic was applied to
synchronize image recording of the high speed camera with the zero
crossing of the ac component of the applied voltage. Thus the
temporal behaviour could be monitored by taking 20--4000 snapshots
(depending on $f$) within a driving period. Measurements
were computer controlled using LabView.

Digital processing of the recorded images allowed
determination of the wave vector $\mathbf{q}$ of the pattern
(using two-dimensional FFT) as well as of the pattern contrast $\Psi$
which we defined as the mean square deviation of the intensity,
$\Psi=\langle (I_{ij}-\langle I_{ij} \rangle )^2 \rangle$ (here
$I_{ij}$ is the intensity of a pixel, and $\langle\rangle$ denotes
averaging over the whole image). This definition satisfactorily discriminates the initial (undistorted) and the patterned state, though it does not
allow distinguishing between various pattern morphologies.

For the precise measurements of the sample impedance, a dielectric
analyser (Novocontrol Alpha equipped with a ZG4 test interface)
was also employed. In order to obtain the values of the anisotropy
of the dielectric permittivity and the electrical conductivity,
the thermostated sample (a commercial cell of $d=50$ $\mu$m) was
put into an electromagnet with the magnetic induction $\mathbf{B}$
perpendicular to the confining electrodes, i.e., perpendicular to
the initial planar director orientation $\mathbf{n}_0$. At $B=0$,
the components $\varepsilon_{\bot}$ and $\sigma_{\bot}$
(perpendicular to $\mathbf{n}$) could be obtained. As the compound
has a positive magnetic susceptibility anisotropy
$\chi_{\mathrm{a}} = \chi_{\|} - \chi_{\bot} > 0$, increasing the
magnetic field above $B_{\mathrm{F}}=(\pi/d)(\mu_0
K_1/\chi_{\mathrm{a}})^{1/2}$ a Freedericksz transition is induced
resulting in a quasi-homeotropic state with the director
orientation nearly parallel to the magnetic field at high
$B=B_{\mathrm{max}} \approx 1 ~\textrm{T}$, thus allowing to
estimate the components $\varepsilon_{\|}$ and $\sigma_{\|}$
(parallel to $\mathbf{n}$). Here $\mu_0$ is the vacuum
permeability and $K_1$ is the splay elastic constant.

Unfortunately, due to $\varepsilon_{\mathrm{a}}<0$, the dc bias
voltage $U_{\mathrm{dc}}$ has a dielectric stabilizing effect
acting against the magnetic field, hence increasing the magnetic
Freedericksz threshold according to
$B_{\mathrm{F}}(U_{\mathrm{dc}})=B_{\mathrm{F}} \,
[1+(U_{\mathrm{dc}}/U_{\mathrm{F}})^2]^{1/2}$, where
$U_{\mathrm{F}}=\pi[K_1/(\varepsilon_0 | \varepsilon_{\mathrm{a}}
|)]^{1/2}$. In order to keep $B_{\mathrm{max}}\gg
B_{\mathrm{F}}(U_{\mathrm{dc}})$ for large dc bias voltage, a
cell with a thickness of $d\approx 1$ mm was constructed. In
this cell, the polyimide coating on the ITO electrodes was
unrubbed, as at such a thickness the surface interactions cannot
provide a uniform orientation in the bulk. Instead, during the
impedance measurements the magnetic field was kept switched on
continuously at $B=B_{\mathrm{\mathrm{max}}}$ and the cell was
rotated alternately to the positions with $\mathbf{B}$ parallel to
the substrates ($\varepsilon_{\bot}$ and $\sigma_{\bot}$ are
measurable) and $\mathbf{B}$ perpendicular to the substrates
(yielding $\varepsilon_{\parallel}$ and $\sigma_{\parallel}$).

\section{\label{sec:results} \bf{Results} }
The planarly aligned 1OO8 exhibits flexodomains as a first instability at pure dc driving, while it shows EC patterns when driven with ac voltage \cite{Salamon2013}. In Section \ref{sec:1OO8} we
explore how superposition of dc and ac voltages alters the pattern
morphologies, the onset characteristics ($U_{\mathrm{c}}$ and $\mathbf{q}_{\mathrm{c}}$)
and their frequency dependence, and up to what combination of the
superposed voltages the uniform state can prevail. In Section
\ref{sec:anticipation} our theoretical considerations are
presented in comparison with the experimental findings. Finally we report in Section
\ref{sec:current} on the results of the conductivity measurements, that are essential for understanding the phenomena under scope.

\subsection{\label{sec:1OO8} Pattern morphologies
and the stability limit curve}

\subsubsection{\label{sec:acdc} Superposition of ac and dc voltages}

\begin{figure}[!h]
\begin{center}
\includegraphics[width=8cm]{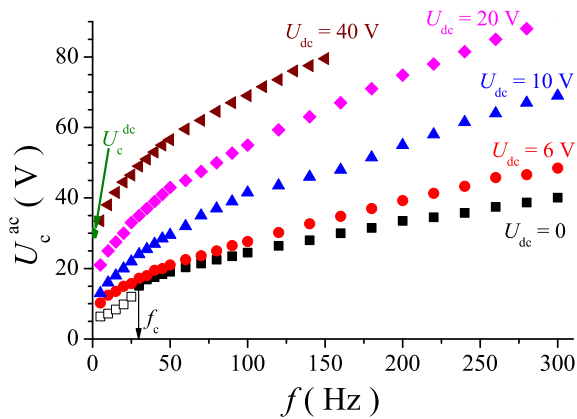}
\end{center}
\caption{(Color online) The frequency dependence of the
ac threshold voltage $U_{\mathrm{c}}^{\mathrm{ac}}$ of EC in 1OO8 at various
dc bias voltages $U_{\mathrm{dc}}$ in the frequency range of 5 Hz $\leq f\leq 300$ Hz. The vertical black arrow indicates the crossover frequency $f_{\mathrm{c}}$ between conductive (opel symbols) and dielectric (solid symbols) EC regimes for $U_{\mathrm{dc}}=0$. At $U_{\mathrm{dc}}\neq 0$ only dielectric EC is observable.} \label{fig:1oo8_Vac(f,Vdc)}
\end{figure}

Figure \ref{fig:1oo8_Vac(f,Vdc)} presents the frequency dependence
of the ac threshold voltages $U_{\mathrm{c}}^{\mathrm{ac}}(f)$ of a $d=19.5$ $\mu$m thick cell, for pure ac ($U_{\mathrm{dc}}=0$)
as well as for different dc bias voltages up to $U_{\mathrm{dc}}=40$ V. For
pure ac driving one sees the usual scenario of standard EC:
conductive regime at low $f$ and dielectric regime above the
crossover frequency $f_{\mathrm{c}} \sim 30$ Hz (see the vertical black arrow in Fig.~\ref{fig:1oo8_Vac(f,Vdc)}).

Note that since 1OO8 exhibits flexodomains for pure dc voltage driving at $U > U_{\mathrm{c}}^{\mathrm{dc}} \approx 30$~V (green arrow in Fig.~\ref{fig:1oo8_Vac(f,Vdc)}), there should exist a transition from the conductive EC roll pattern to FD as the ac frequency is reduced. This crossover
occurs, however, at ultralow $f$ at a few mHz \cite{Salamon2013},
therefore it is not shown in Fig.~\ref{fig:1oo8_Vac(f,Vdc)}.

\begin{figure}[!h]
\begin{center}
 \includegraphics[width=8cm]{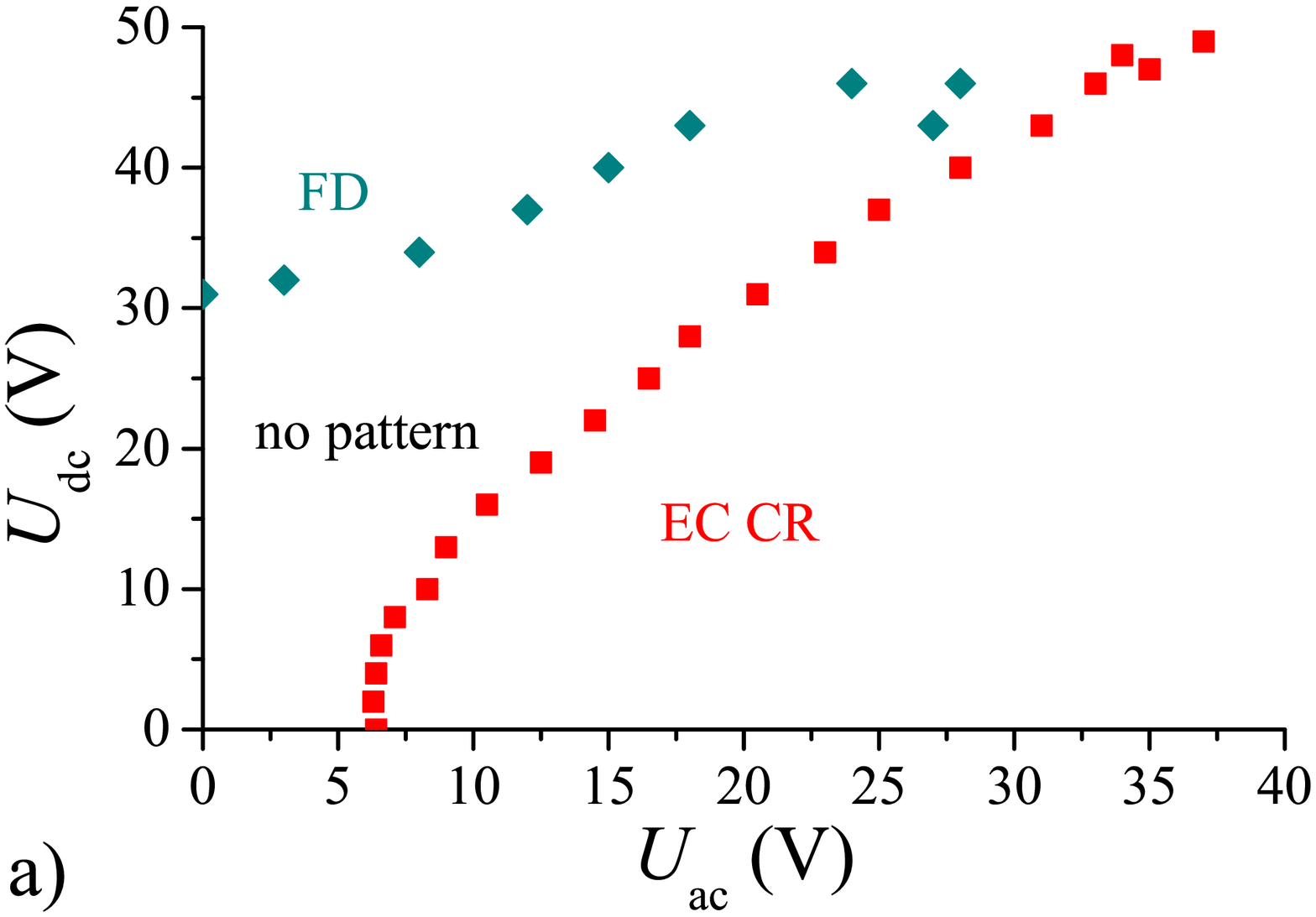}\\ 
 \includegraphics[width=8cm]{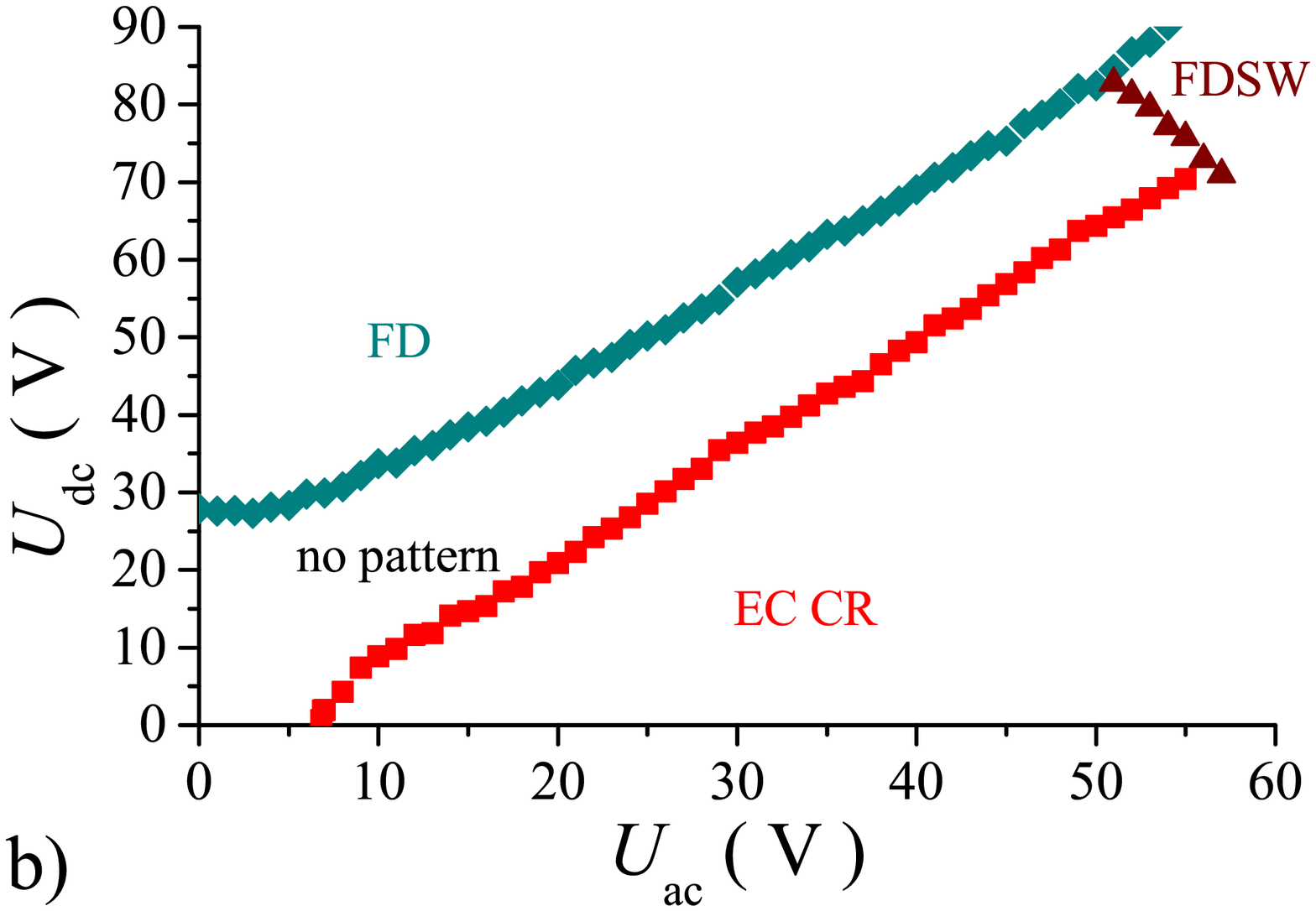}\\ 
 \includegraphics[width=8cm]{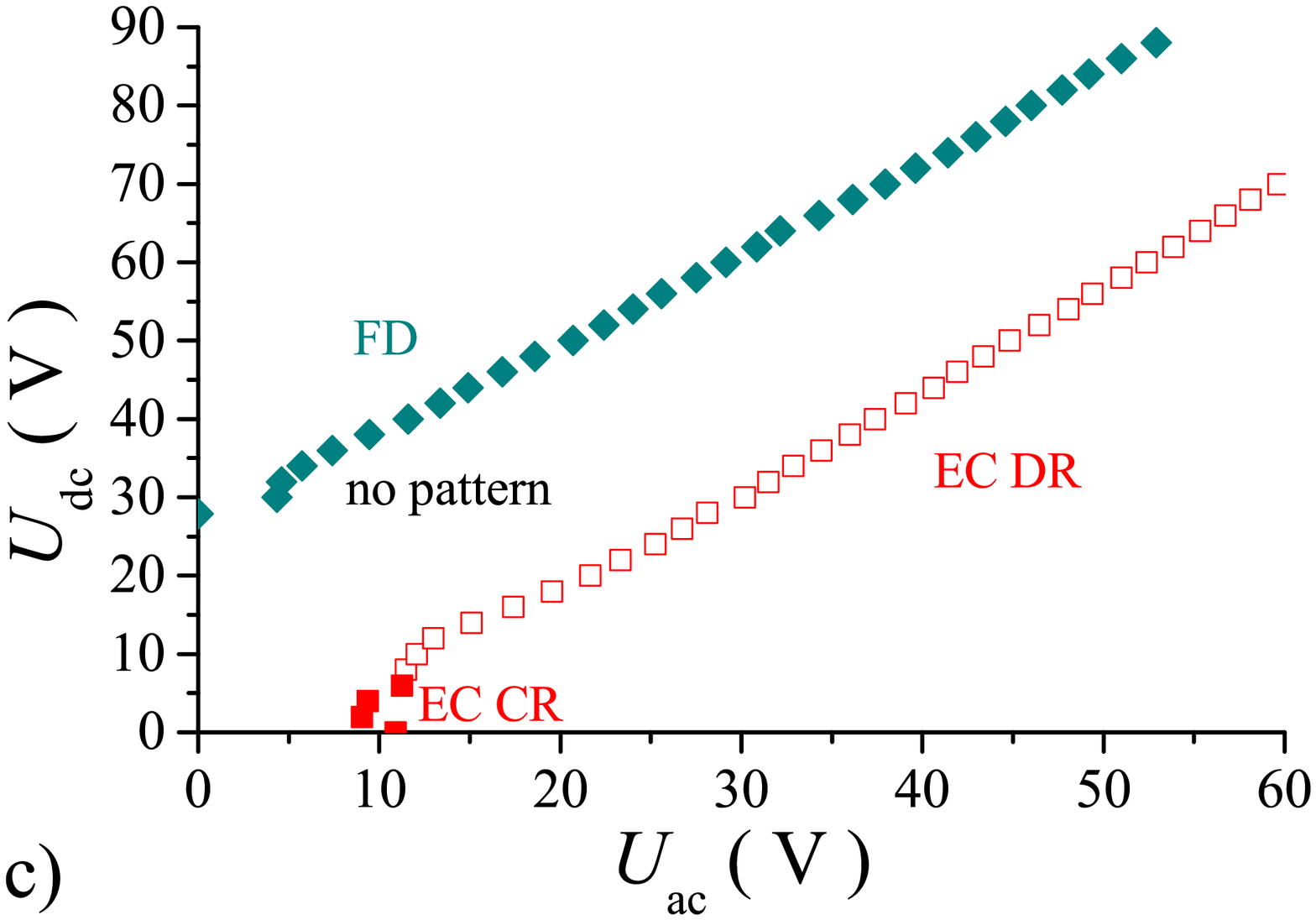}  
\end{center}
\caption{(Color online) Morphological phase diagram under combined dc and ac voltages with a) $f=2$ Hz, b) $f=5$ Hz, and c) $f=10$ Hz. The abbreviations EC CR, EC DR and FDSW correspond to conductive EC regime, dielectric EC regime and flexodomains of short wavelength, respectively.} \label{fig:1oo8_5Hz}
\end{figure}

Applying a small dc bias
voltage, the crossover frequency is substantially reduced; for
$U_{\mathrm{dc}}\gtrsim 4$ V only the dielectric regime becomes detectable
for $f>5$ Hz. At a fixed frequency, the ac threshold voltage $U_{\mathrm{c}}^{\mathrm{ac}}$ increases
monotonically with the dc bias voltage.

In Figs.~\ref{fig:1oo8_5Hz}(a)--\ref{fig:1oo8_5Hz}(c), we
present the morphological phase diagrams (the SLC in the
$U_{\mathrm{ac}}$--$U_{\mathrm{dc}}$ plane) of the same cell for $f=2$ Hz, $f=5$ Hz and $f=10$ Hz, respectively.
Though these figures resemble the ones obtained previously for
Phase 5 (e.g. Figs.~1 and 2 of \cite{Salamon2014}), we have to
emphasize an important difference: the upper branch of the SLC
here corresponds to the threshold of flexodomains (instead of the
dc mode of EC).

\begin{figure}[!ht]
\begin{center}
\includegraphics[width=8cm]{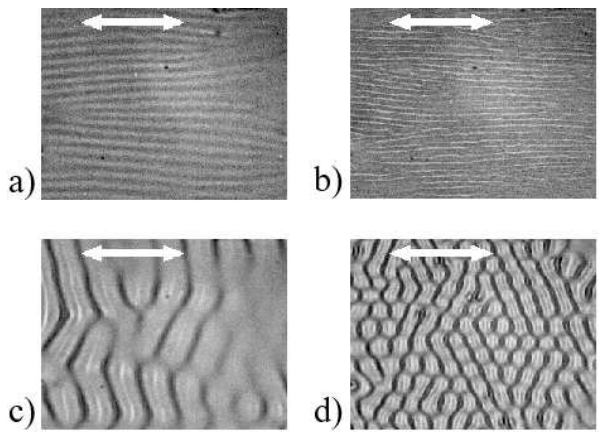}
\end{center}
\caption{Typical pattern morphologies under combined dc and ac voltage with $f=5$
Hz: a) flexodomains at $U_{ac}=1$ V, $U_{dc}=30$ V; b)
flexodomains at $U_{ac}=48$ V, $U_{dc}=80$ V; c) oblique rolls in the conductive EC regime at $U_{ac}=7.4$ V, $U_{dc}=0$ V; d) oblique rolls in the conductive EC regime at $U_{ac}=60$ V, $U_{dc}=72$ V. The double arrows show the
direction of the initial director; their length corresponds to 100
$\mu$m ($d=19.5$ $\mu$m).} \label{fig:Image_5Hz}
\end{figure}

It can be seen that adding an ac voltage results in an increase of the dc FD threshold $U_{\mathrm{dc}}$ which
becomes nearly linear for higher $U_{\mathrm{ac}}$. This behaviour is in accordance with previous findings on FD in other
nematics \cite{Marinov2006,Tadapatri2012}. Moving along this
branch the type of the pattern (FD) remains unaltered [see Figs.~\ref{fig:Image_5Hz}(a) and \ref{fig:Image_5Hz}(b)], only a
decrease of the pattern wavelength can be observed.  The lower branch of the SLC corresponds to
conductive EC oblique rolls [see Figs.~\ref{fig:Image_5Hz}(c) and \ref{fig:Image_5Hz}(d)]. It is seen
that, in agreement with Fig.~\ref{fig:1oo8_Vac(f,Vdc)}, increasing
the dc bias voltage results in higher ac thresholds and also in some
reduction of the pattern wavelength.

For low frequencies of the ac voltage component ($f=2$ Hz, Fig.~\ref{fig:1oo8_5Hz}(a)), the two SLC branches
join directly, representing a crossover from conductive EC to FD patterns at
$U_{\mathrm{ac}} \sim 30$ V and $U_{\mathrm{dc}} \sim 45$ V.

\begin{figure}[!h]
\begin{center}
\includegraphics[width=8cm]{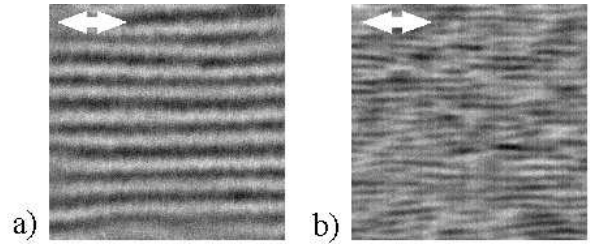}
\end{center}
\caption{Two types of flexodomains under combined dc and ac voltage with $f=5$ Hz:
a) FD at $U_{ac}=12$ V, $U_{dc}=40$ V; b) FDSW at $U_{ac}=55$ V,
$U_{dc}=80$ V. Note the huge difference in the wavelength of the
pattern. The double arrows show the initial director orientation; their length corresponds to 20 $\mu$m ($d=19.5$ $\mu$m).}
\label{fig:Image_FD2}
\end{figure}

At $f=5$ Hz (Fig.~\ref{fig:1oo8_5Hz}(b)) the upper (FD) and lower (conductive EC) branches run almost parallel at higher voltages. In this case the SLC has a third, connecting branch which limits
the pattern-free region. The pattern appearing here corresponds
also to parallel stripes, but with a much lower contrast and much
shorter wavelength than those of the FD along the upper branch; it
is identified as another type of flexodomains (denoted as FDSW).
As illustrations, Figs.~\ref{fig:Image_FD2}(a) and \ref{fig:Image_FD2}(b) show
snapshots of the FD and the FDSW patterns, respectively, at the
same (higher) magnification after contrast enhancement. Outside the SLC, near to the crossing of the branches, there is a voltage range where
FDSW may coexist with the usual FD. The two kinds of patterns occupy different locations; while FD is stationary, FDSW fluctuates with the driving voltage. Similarly, near the crossing of the EC and FDSW branches, FDSW may coexist with conductive EC rolls; in this case the patterns emerge at the same location, but in different time windows within the same driving period of the ac voltage (similarly to the ultralow $f$ behaviour of calamitic materials \cite{Salamon2013}).

\begin{figure}[!h]
\begin{center}
 \includegraphics[width=8cm]{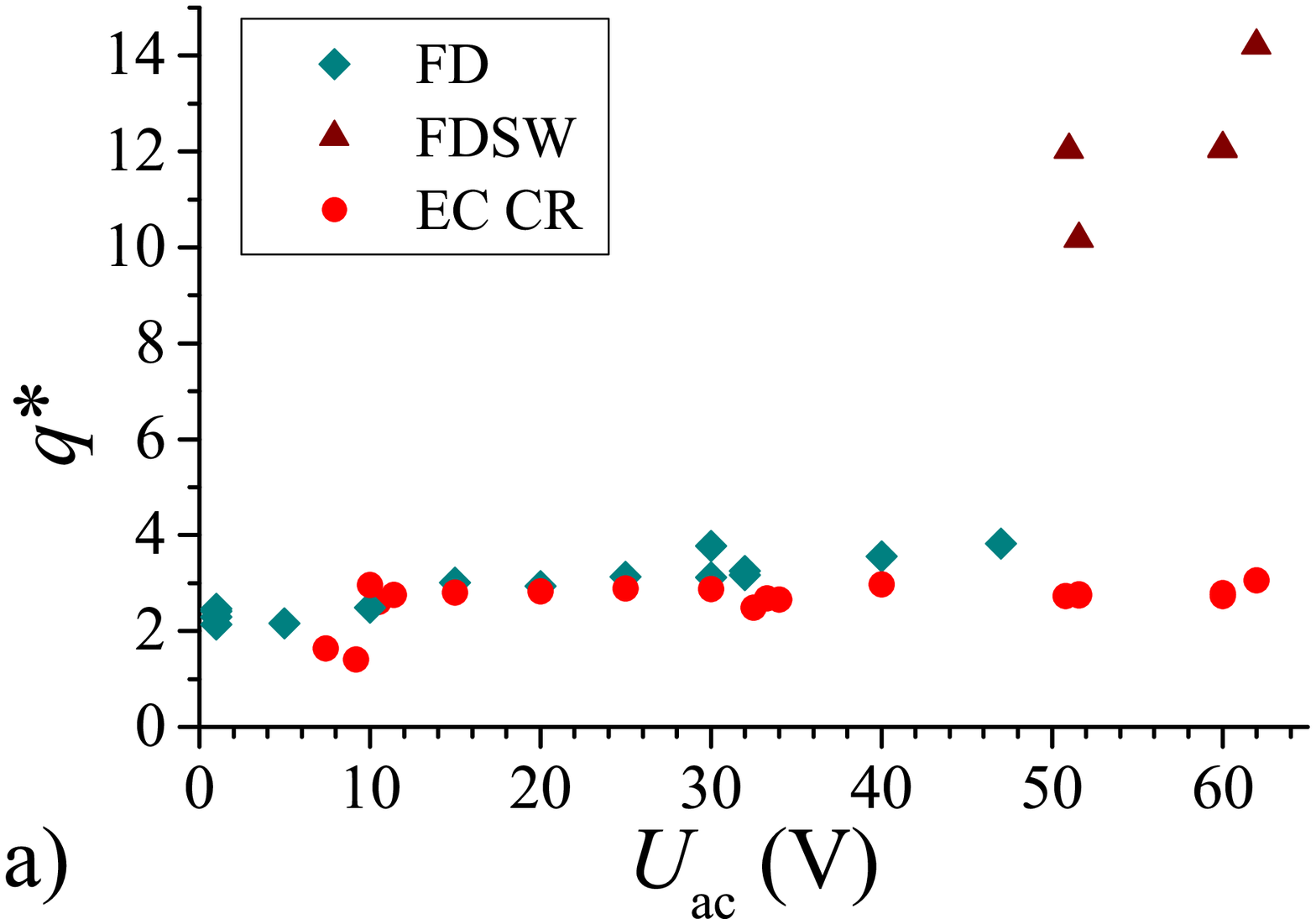}\\ 
 \includegraphics[width=8cm]{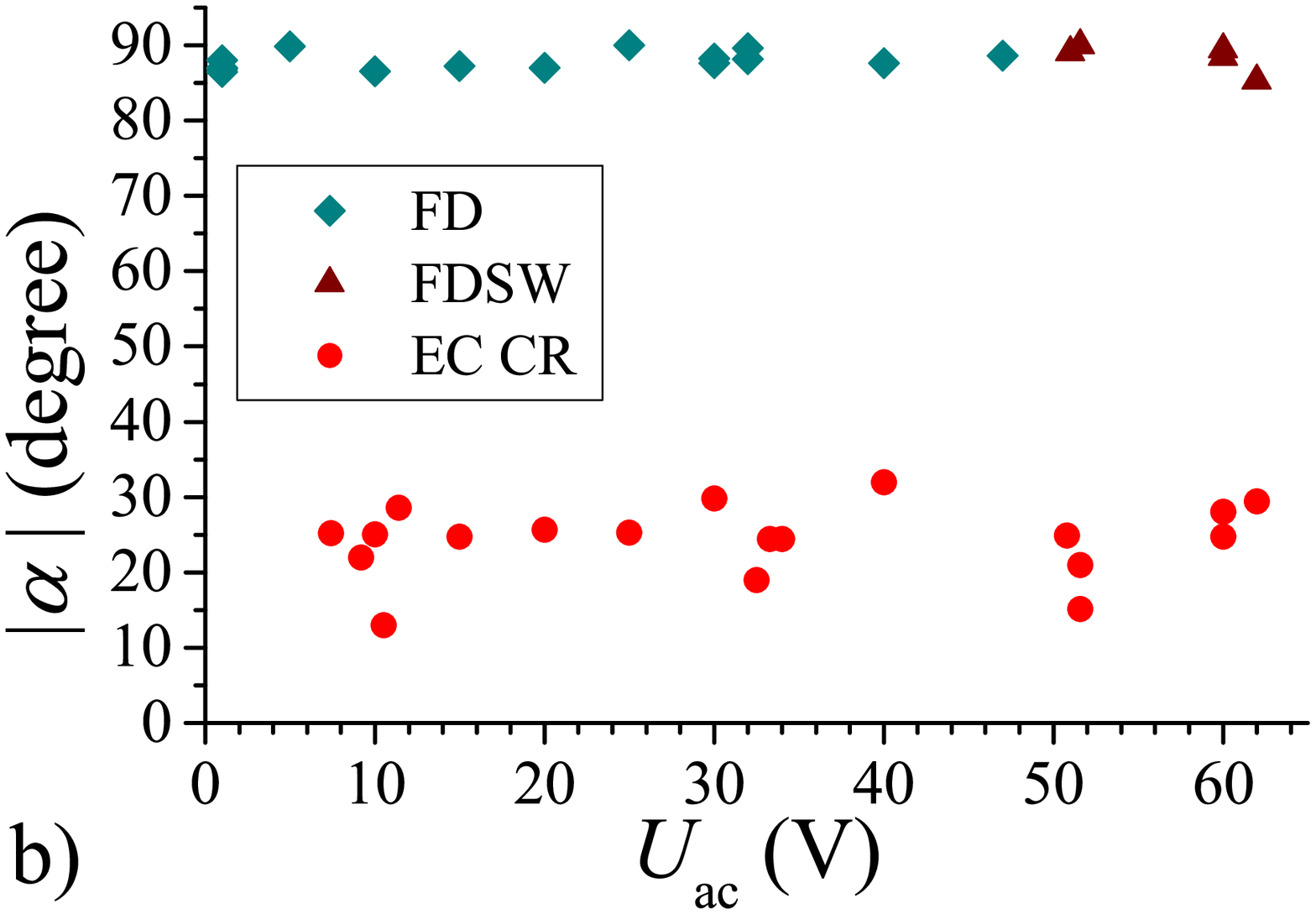}\\ 
\end{center}
\caption{(Color online) Dependence of a) the dimensionless wavenumber $q^*$, and b) the obliqueness angle $|\alpha|$ of flexodomains and EC patterns on the ac voltage at $f=5$ Hz, close to threshold.} \label{fig:1oo8_q}
\end{figure}

The change of the dimensionless wave number $q^{*} = q d/ \pi = 2 d/\lambda$ ($\lambda$ is the wavelength of the pattern) and the obliqueness angle $\alpha$ at threshold (along the SLC) is depicted in Figs.~\ref{fig:1oo8_q}(a) and \ref{fig:1oo8_q}(b), respectively. It can be seen that $q^{*}$ of FD and of the conductive regime of EC are very similar ($q^{*}\sim 1.5$--3.5) and depend only slightly on $U_{\mathrm{ac}}$, while for FDSW one finds much larger values ($q^{*}\sim 10$--15) with a stronger $U_{\mathrm{ac}}$ dependence. In contrast to $q^{*}$, the obliqueness angle is not affected by the FD--FDSW transition; both are parallel to $\mathbf{n}_0$. Note that for both kinds of flexodomains the apparent periodicity [the distance between subsequent dark stripes in Figs.~\ref{fig:Image_5Hz}(a)--(b) and \ref{fig:Image_FD2}(a)--(b)] corresponds to the half wavelength, while for EC in
Figs.~\ref{fig:Image_5Hz}(c)--(d) black lines are repeated with the periodicity of the full wavelength.

For a slightly higher frequency [$f=10$ Hz, Fig.~\ref{fig:1oo8_5Hz}(c)], the morphological diagram is similar: the
upper and lower branches of the SLC run nearly parallel. There is,
however, a transition in the pattern type along the lower (EC)
branch of the SLC. The conductive oblique rolls are seen only for
$U_{\mathrm{dc}}\leq 6$ V; for higher dc bias voltage the pattern switches
to dielectric rolls yielding a jump in the wavelength of the
pattern and also in the obliqueness (the dielectric rolls are
normal to $\mathbf{n}_0$ along the SLC). Another important difference is that at
this $f$, the two branches of the SLC are not connected within the
experimentally accessible voltage range. The inhibition of pattern formation by
superposing ac and dc voltages is so effective  that no pattern
appears at voltages exceeding several times the individual
thresholds, e.g. at $U_{\mathrm{dc}} \approx 90$ V combined with
$U_{\mathrm{ac}} \approx 60$ V.

For  even higher frequencies the SLC looks similar to that in Fig.~\ref{fig:1oo8_5Hz}(c); except that the EC branch is shifted towards higher
ac voltage and for $f\gtrsim 30$ Hz only dielectric rolls are observable.

\subsubsection{\label{sec:acac} Superposition of ac voltages of different frequencies}

Besides  exploring the behaviour at superposed ac and dc voltages,
we have also tested what happens if two ac voltages ($U_{\mathrm{ac1}}$  and
$U_{\mathrm{ac2}}$) with substantially different frequencies ($f_1 \gg f_2$,
respectively) are superposed. To have both frequencies in the conductive regime of EC, the crossover frequency $f_{\mathrm{c}}$ had to be pushed above $f_1$, which could be realized using thicker cells than in Section \ref{sec:acdc}.

\begin{figure}[!h]
\begin{center}
\includegraphics[width=8cm]{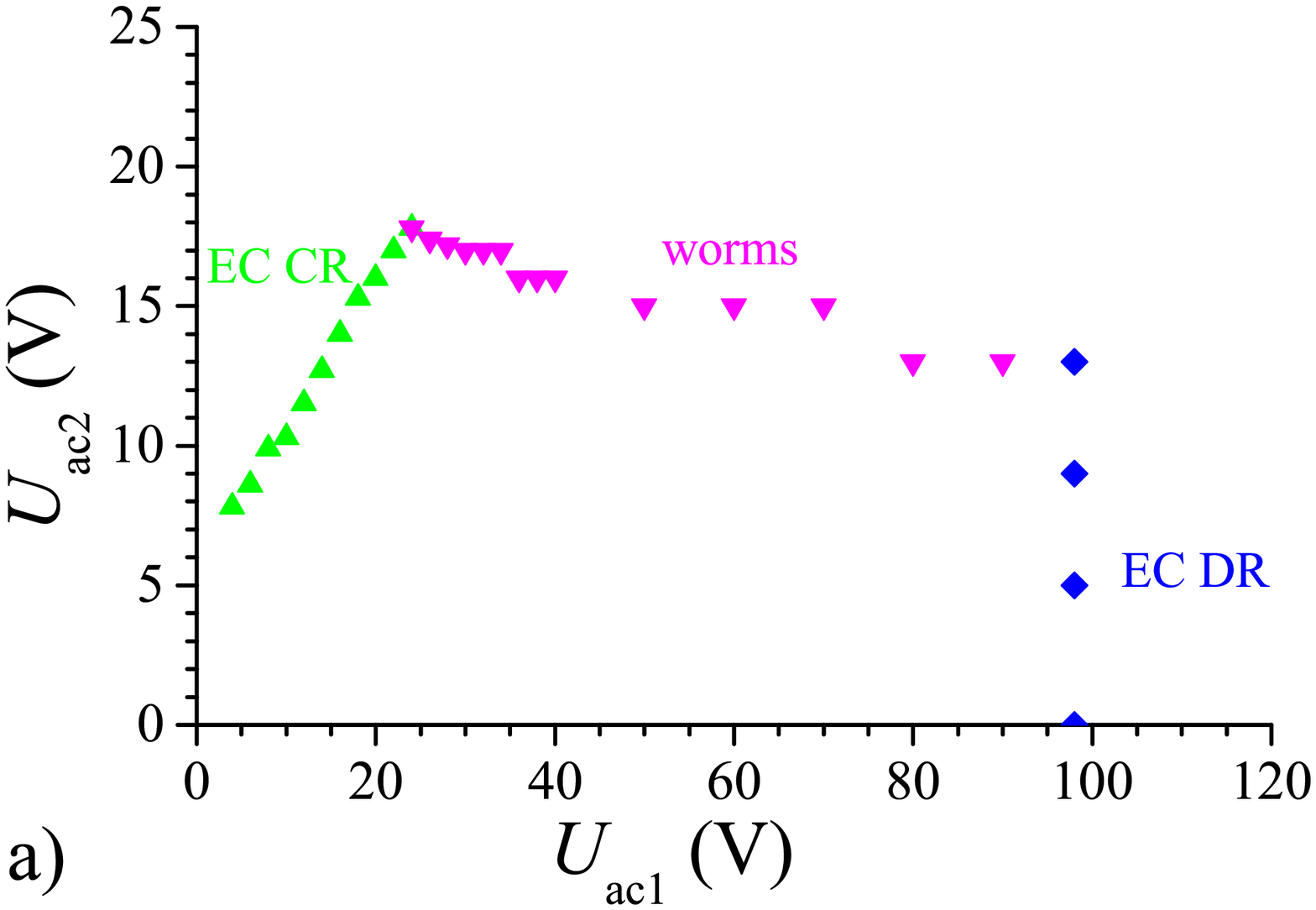}\\ 
\includegraphics[width=8cm]{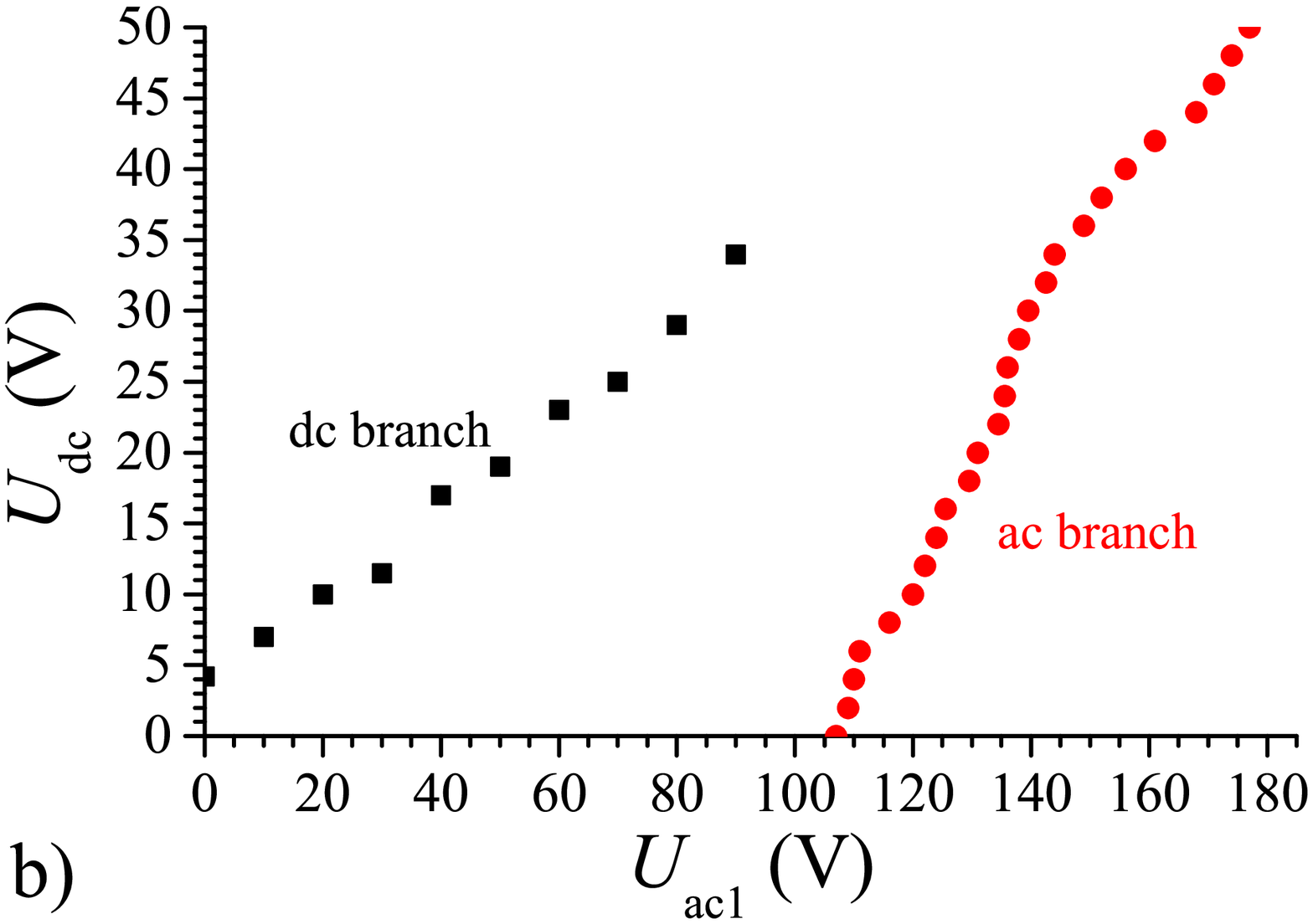}  
\end{center}
\caption{(Color online) a) Morphological phase diagram at
superposing an ac voltage $U_{\mathrm{ac1}}$ of $f_1 = 400$ Hz (horizontal
axis) with another ac voltage $U_{\mathrm{ac2}}$ of $f_2 = 10$ Hz ($U_2$,
vertical axis). b) The stability limiting curves for the case when the ac voltage $U_{\mathrm{ac2}}$ of $f_2 = 10$ Hz is replaced by a dc voltage $U_{\mathrm{dc}}$. } \label{fig:ac10_ac400}
\end{figure}

\begin{figure}[!h]
\begin{center}
\includegraphics[width=8cm]{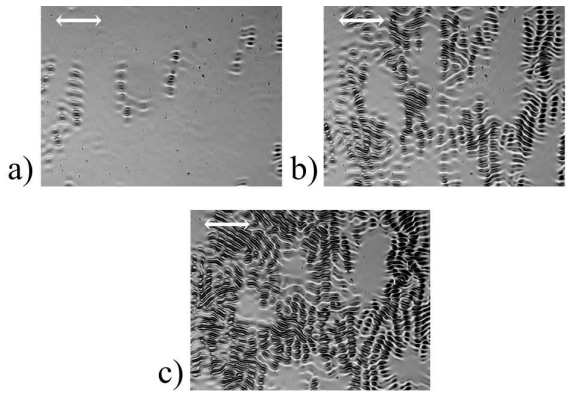}
\end{center}
\caption{Localized electroconvection structures and their development to extended pattern at the superposition of two ac voltages.
The high frequency component was kept constant, while the low $f$ component was increased from (a) around threshold to (b) 4\% above threshold and (c) 8\% above threshold.
The double arrow shows the initial director orientation; its length corresponds to 400 $\mu$m ($d=50$ $\mu$m).}
\label{fig:Image_worm}
\end{figure}

Figure \ref{fig:ac10_ac400}(a) exhibits
the morphological phase diagram of a $d=50$ $\mu$m thick sample in
the $U_{\mathrm{ac1}}$--$U_{\mathrm{ac2}}$ plane for $f_1=400$ Hz and $f_2=10$ Hz. The SLC has
three branches. Along the low frequency ($f_2$) branch, oblique
rolls of conductive EC are seen and the threshold voltage is
increasing upon superposing a voltage with high frequency. The
high frequency ($f_1$) branch of the SLC is practically vertical;
the threshold voltage of the dielectric EC occurring here is not
altered by adding a lower frequency component. The pattern free
region is closed by the third SLC branch running across the
previous two ones, indicating sharp morphological transitions at
the intersection points. Indeed, along this branch, the pattern is
very different; instead of extended rolls, the pattern is rather
localized to small regions in space, known as worms \cite{Dennin1996,Tu1997,Riecke1998,Bisang1999}. A
sample snapshot is shown in Fig.~\ref{fig:Image_worm}(a). Increasing the low frequency voltage component $U_2$ while keeping $U_1$ constant, these worms serve as seeds for gradually extending the roll pattern to the full viewed area. This process is illustrated in Figs.~\ref{fig:Image_worm}(b) and \ref{fig:Image_worm}(c). The slope
of the $U_{\mathrm{ac2}}(U_{\mathrm{ac1}})$ threshold curve is slightly negative in this branch; adding a
high $f$ voltage reduces the low $f$ threshold.

We note that a similar worm state has been reported in \cite{Pietschmann2010}, for the case when $f_2$ was chosen in the conductive regime and $f_1=2 f_2$ was above $f_{\mathrm{c}}$ in the dielectric regime. This may imply that the condition $f_2 < f_{\mathrm{c}} < f_1$ is more important in inducing the worm scenario, than the actual value of the $f_1 : f_2$ ratio.

For comparison,
the SLC for the superposition of ac and dc voltages is shown
in Fig.~\ref{fig:ac10_ac400}(b). In this case, the two branches of the
SLC let the pattern-free region extend to high $U_{\mathrm{ac1}}$, $U_{\mathrm{dc}}$ voltage
combinations, similarly to the case shown in Fig.~\ref{fig:1oo8_5Hz}(c). These observations prove that
the effective pattern inhibition is really a consequence of the applied
dc voltage. Noticeably in Fig.~\ref{fig:ac10_ac400}(b), the
ac branch (EC DR) of the SLC is shifted to much higher ac voltage compared to the cases in Fig.~\ref{fig:1oo8_5Hz}, due to
the higher driving frequency, while the dc branch (EC CR) of the SLC
appears at much lower dc voltage, as in this thicker sample EC
occurs at dc driving instead of FD.

\subsection{\label{sec:anticipation} Theoretical considerations}

As already
mentioned in Section \ref{sec:intro}, by now the theoretical
description of electroconvection and of flexodomains (the extended
SM) has been worked out for pure ac or dc drivings
\cite{Krekhov2011,Krekhov2008}. Since the patterns emerge
continuously from the initial state at onset (corresponding to
a forward bifurcation), the onset characteristics ($U_{\mathrm{c}}$ and
$\mathbf{q}_{\mathrm{c}}$) can be obtained from a linear stability analysis
of the underlying nemato-hydrodynamic equations.
The linearized equations and the details of the numerical procedure are presented in \cite{Krekhov2008}.
In case of combined ac+dc driving the code used in \cite{Krekhov2008} has been modified by including the additional dc voltage (see \cite{Krekhov2014} for some examples of the onset calculations).

To compare the experimental results with the theoretical calculations, the following material parameter set of 1OO8 has been used;
elastic constants: $K_{11} = 7.248$ pN, $K_{22} = 3.67$ pN, $K_{33} = 9.379$ pN \cite{Salamon2013}; viscosity coefficients (in lack of measurements MBBA values were taken): $\alpha_1 = -18.1$ mPa s, $\alpha_2 = -110.4$ mPa s, $\alpha_3 = -1.1$ mPa s, $\alpha_4 = 82.6$ mPa s, $\alpha_5 = 77.9$ mPa s, $\alpha_6 = -33.6$ mPa s; dielectric permittivities: $\varepsilon_{\bot} = 5.007$, $\varepsilon_{\mathrm{a}} = - 0.488$; electrical conductivities: $\sigma_{\bot} = 3$ nS m$^{-1}$, $\sigma_{\mathrm{a}}/\sigma_{\bot} = 0.2$; flexoelectric coefficients: $e_1 = 6.8$ pC m$^{-1}$, $e_3 = 0$ pC m$^{-1}$ and sample thickness: $d = 20$ $\mu$m.

\begin{figure}[!h]
\begin{center}
\includegraphics[width=8cm]{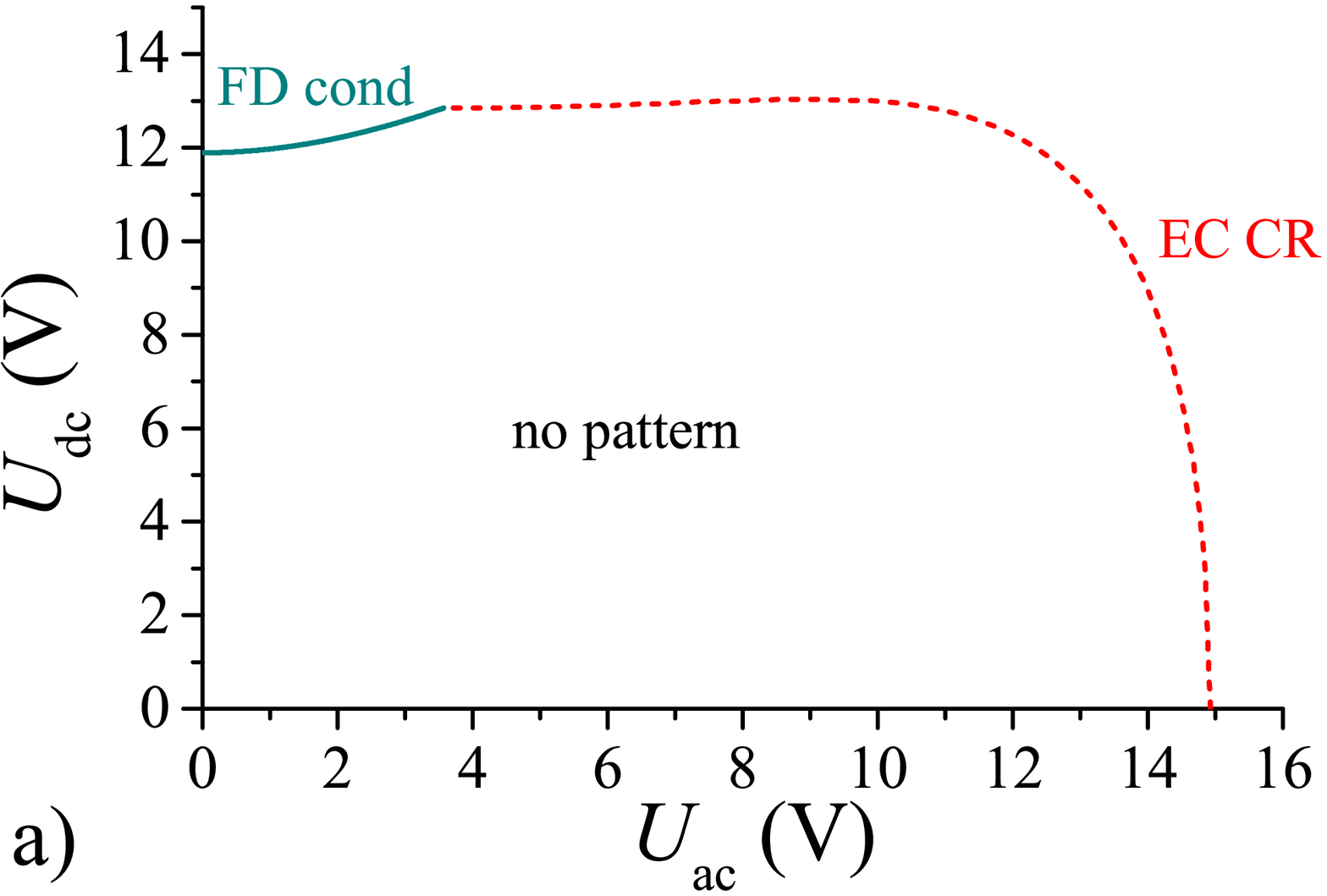}\\ 
\includegraphics[width=8cm]{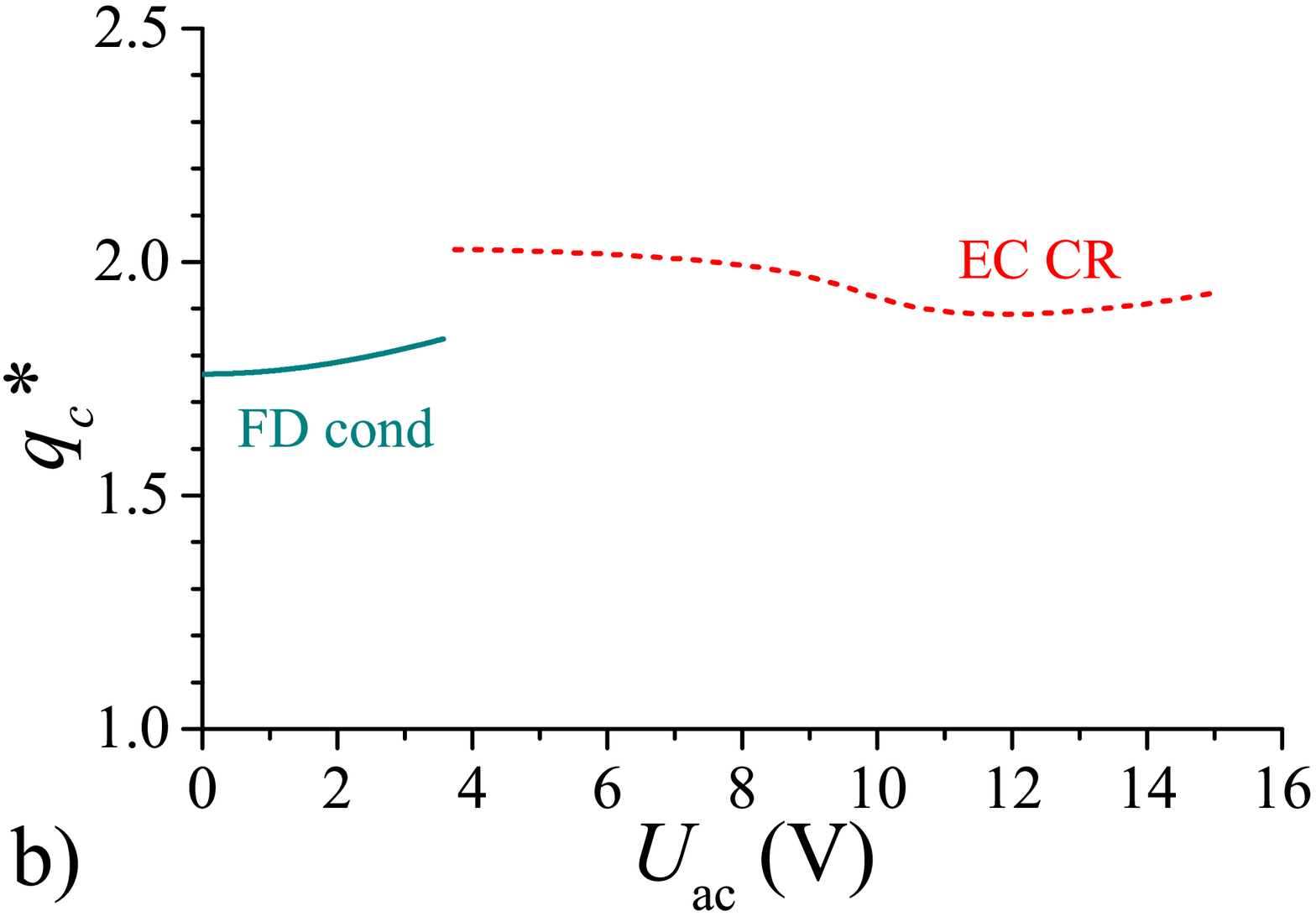}\\ 
\includegraphics[width=8cm]{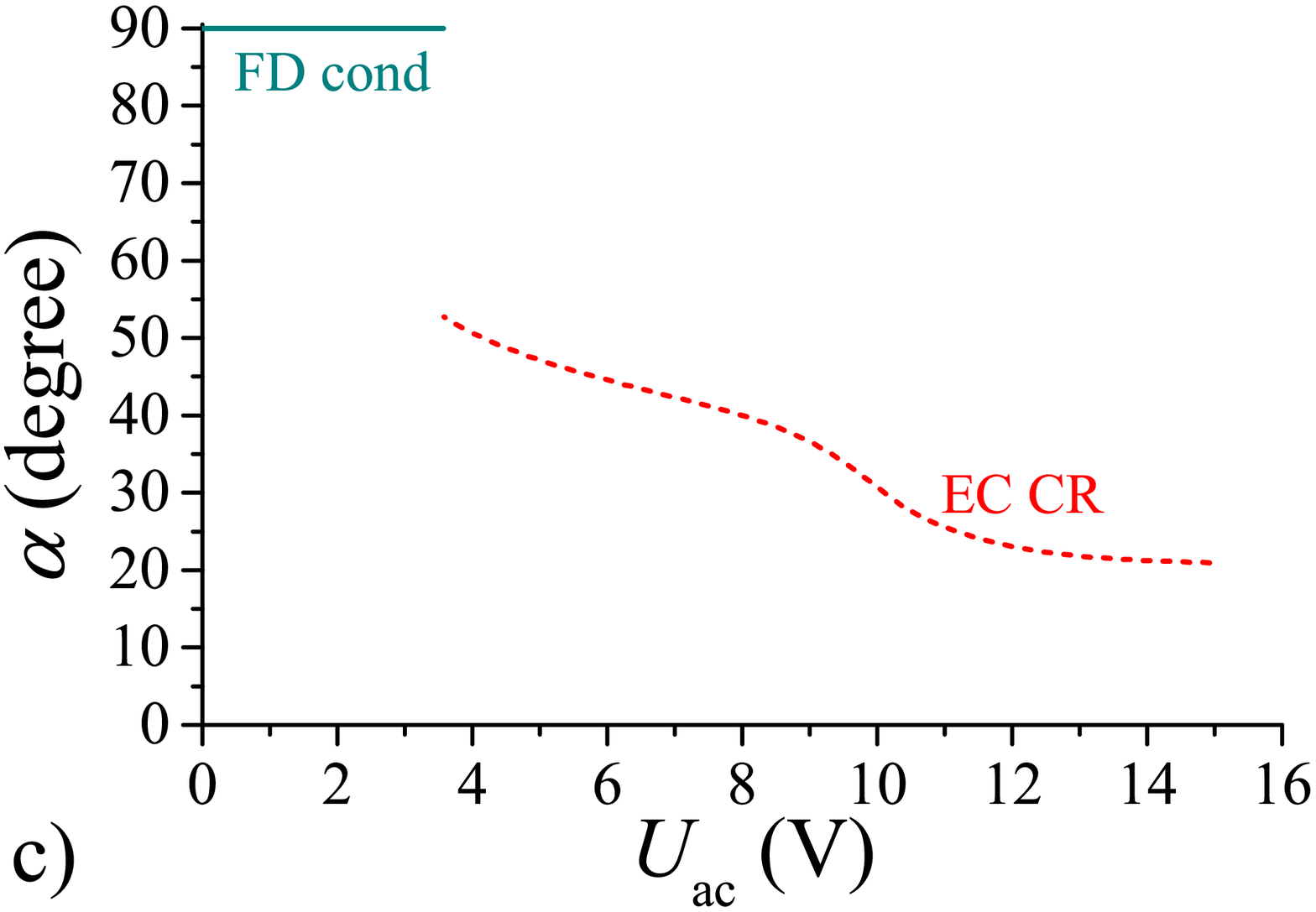} 
\end{center}
\caption{(Color online) Phase diagram under combined dc and ac voltage with $f = 2$~Hz exhibiting flexodomains of conductive type (FD cond) at dc driving and conductive EC regime (EC CR) at ac driving: (a) Stability limiting curve in the $U_{\mathrm{ac}}$--$U_{\mathrm{dc}}$ plane, (b) the critical dimensionless wave numbers $q_{\mathrm{c}}^*$ and (c) the obliqueness angles $\alpha$ along the SLC.} \label{fig:flexcond_calc}
\end{figure}

\begin{figure}[!h]
\begin{center}
\includegraphics[width=8cm]{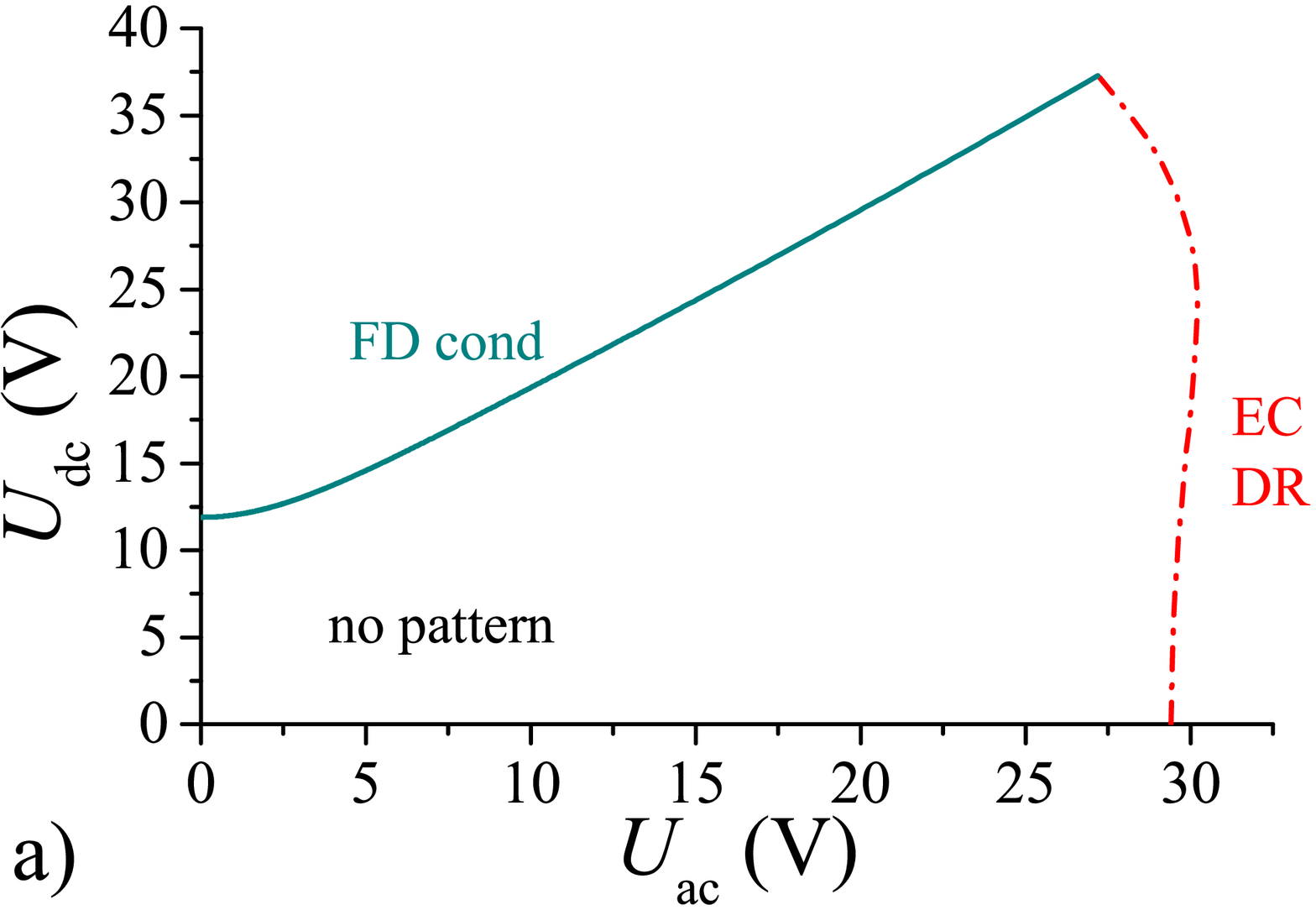}\\ 
\includegraphics[width=8cm]{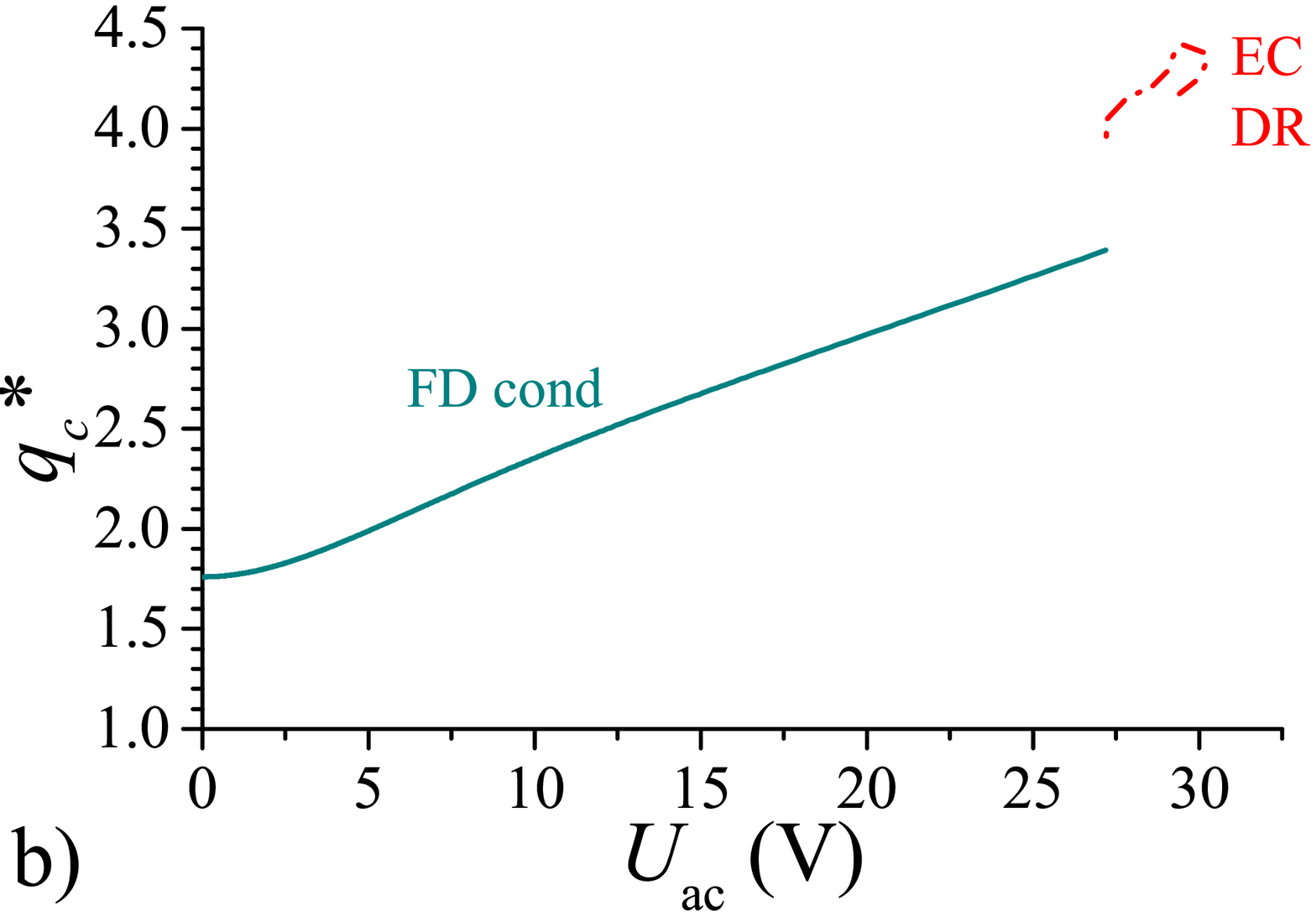} 
\end{center}
\caption{(Color online) Phase diagram under combined dc and ac voltage with $f = 10$~Hz exhibiting flexodomains of conductive type (FD cond) at dc driving and dielectric EC regime (EC DR) at ac driving: (a) Stability limiting curve in the $U_{\mathrm{ac}}$--$U_{\mathrm{dc}}$ plane and (b) the critical dimensionless wave numbers $q_{\mathrm{c}}^*$ along the SLC.} \label{fig:flexdiel_calc}
\end{figure}

The results of the calculations are shown in Figs.~\ref{fig:flexcond_calc} and \ref{fig:flexdiel_calc} for two scenarios; FD occur at dc driving, while at ac driving the EC regime is conductive (low $f$)
or dielectric (high $f$), respectively. The main features of the SLC and the ac voltage dependence of the dimensionless critical wave number $q_c^*$ and the obliqueness angle $\alpha$ strongly resemble those obtained for Phase 5
\cite{Krekhov2014}. Comparing the experimental curves in
Figs.~\ref{fig:1oo8_5Hz}(a), \ref{fig:1oo8_5Hz}(b) and \ref{fig:1oo8_q} with the calculated ones in Figs.~\ref{fig:flexcond_calc} and \ref{fig:flexdiel_calc}, we can
see that the dc branch of the SLC follows the predictions: adding
an ac component increases the dc threshold voltage as well as the critical wave number of FD. For the ac branch of
the SLC, however, there is a mismatch between the theoretical predictions and
the experimental findings. According to the calculations and also the qualitative analysis presented in \cite{Krekhov2014}, the dc voltage should reduce the ac
threshold of EC in the conductive regime [see Fig.~\ref{fig:flexcond_calc}(a)], and in the dielectric regime the same reduction should occur after a minor initial increase [see Fig.~\ref{fig:flexdiel_calc}(a)]. In contrast to that, in the experiments a substantial increase of the ac threshold upon dc bias was detected (for both conductive and dielectric
EC), similarly to the recent findings on the nematic Phase 5
\cite{Salamon2014}.

The recent theoretical analysis of flexodomains \cite{Krekhov2011} has pointed out that in case of pure ac driving the
equations have two solutions with different time symmetries: the conductive mode implies stationary modulation of the out-of-plane director component, while for the dielectric mode it oscillates with the ac frequency. The threshold voltages and critical
wave numbers of the two solutions are different; which of them has a lower threshold depends on the
flexoelectric coefficients, on the dielectric anisotropy, on the elastic constants and on the
frequency of driving. For typical material parameter sets,
conductive FD are expected for low $f$, but dielectric ones for
high $f$. Nevertheless, the frequency induced transition from
conductive FD to dielectric FD, theoretically predicted in \cite{Krekhov2011}, has not yet been reported, as FD
are usually not observed at high $f$ due to their high threshold voltage,
which is either unreachable experimentally or because EC sets in already at
lower $U_{\mathrm{ac}}$.

Adding a dc bias voltage  breaks the time symmetry of the
equations. However, the solution can be associated with one of the three (dc, conductive, or dielectric) modes according to the temporal dynamics of the out-of-plane director component in leading order.

\begin{figure}[!h]
\begin{center}
\includegraphics[width=8cm]{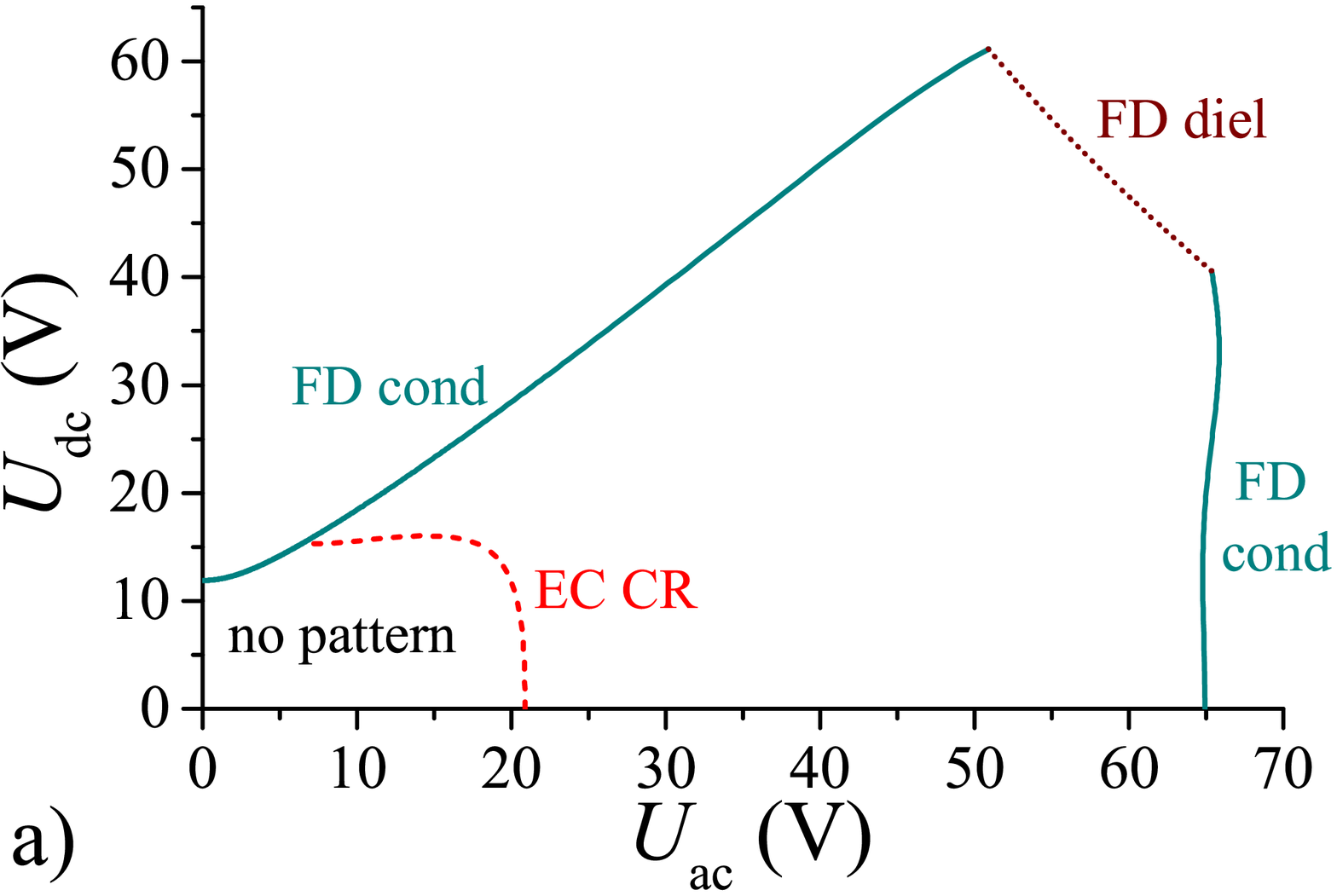}\\ 
\includegraphics[width=8cm]{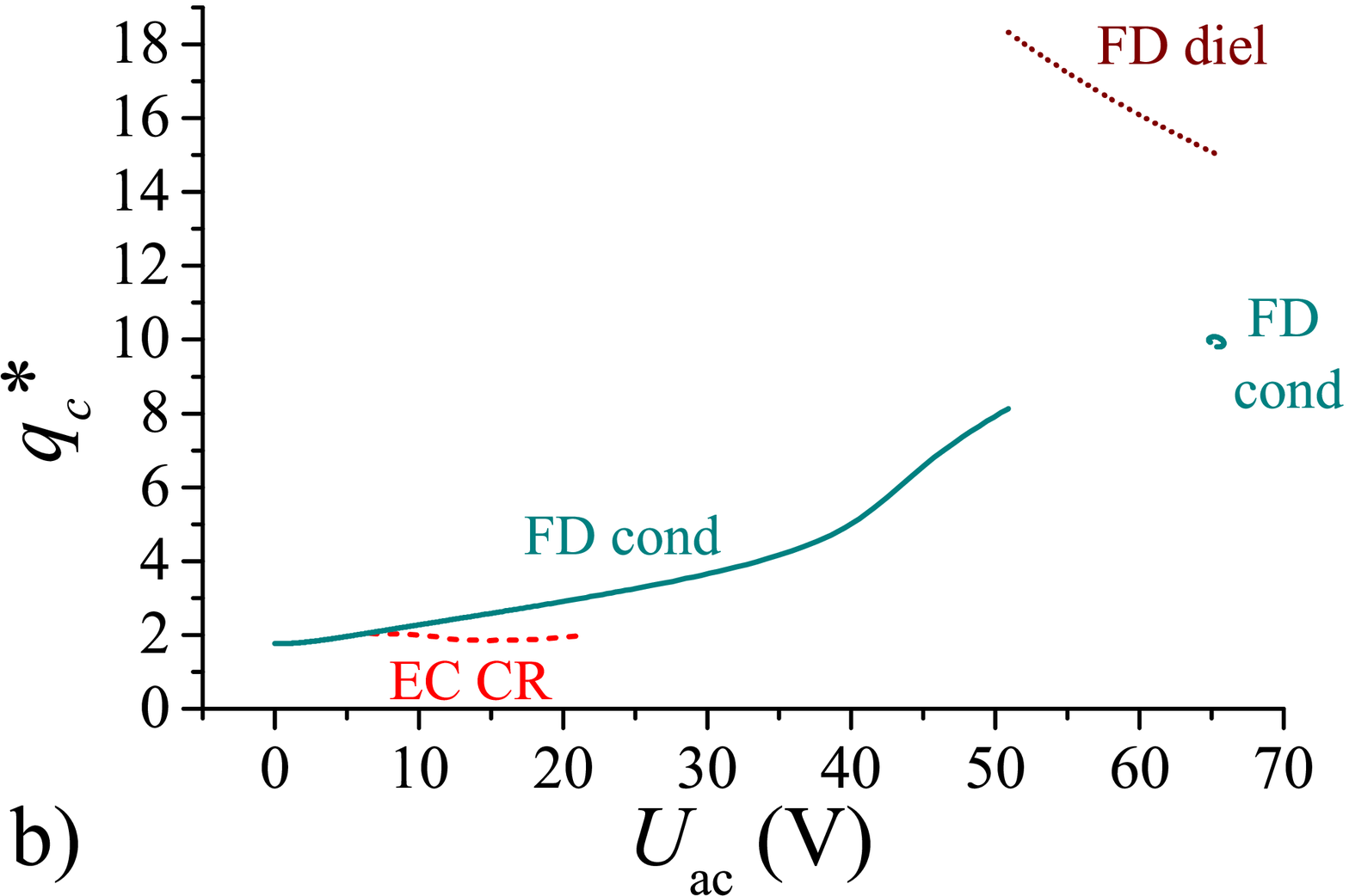}\\ 
\includegraphics[width=8cm]{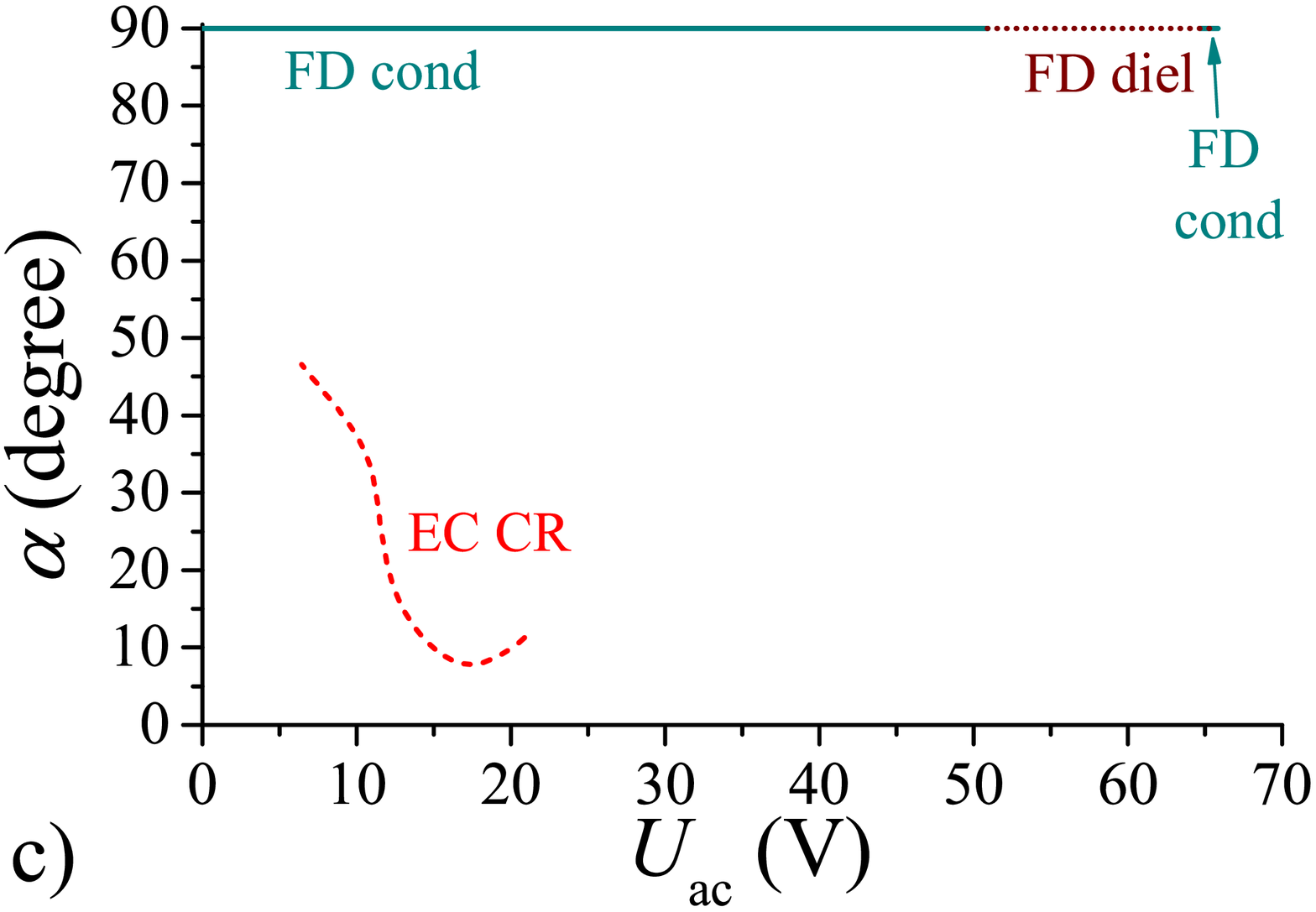} 
\end{center}
\caption{(Color online)
Phase diagram under combined dc and ac voltage with $f = 5$~Hz exhibiting transition between flexodomains of conductive type (FD cond) and of dielectric type (FD diel): (a) Stability limiting curve in the $U_{ac}$-$U_{dc}$ plane, (b) the critical dimensionless wave numbers $q_{\mathrm{c}}^*$ and (c) the obliqueness angles $\alpha$ along the SLC.
For comparison the SLC for the conductive EC regime (EC CR) is also shown with lower threshold voltages than for FD.} \label{fig:Calculated_FD2}
\end{figure}

In order to verify the possibility of the transition from conductive FD to dielectric FD under combined dc and ac driving, their stability diagram has been calculated for 1OO8 at an intermediate frequency $f = 5$~Hz of the ac voltage component. The linear stability analysis of FD demonstrate that indeed, as can be seen in Fig.~\ref{fig:Calculated_FD2}(a), the stability limiting curve has three branches: conductive FD (FDcond) starting from both the dc and the ac axes and a crossing branch of dielectric FD (FDdiel) at high dc and ac voltages.
The wavenumber of FDcond increases with increasing $U_{\mathrm{ac}}$ [see Fig.~\ref{fig:Calculated_FD2}(b)], but has a huge jump at the transition to FDdiel; in the FDdiel range one has a slope of opposite sign. For both kinds of flexodomains the stripes are parallel to $\mathbf{n}_0$ [$\alpha = 90^{\circ}$, see Fig.~\ref{fig:Calculated_FD2}(c)], and the direction of $\mathbf{q}$ does not alter with $U_{\mathrm{ac}}$.

The material parameter studies show that the large elastic anisotropy of 1OO8, $\delta K = (K_{11} - K_{22})/(K_{11} + K_{22}) = 0.33$, is a prerequisite for the transition between two types of FD under combined dc and ac voltages.
In fact, such transition has not been found in the calculations when using the material parameters of the nematic liquid crystal MBBA where $\delta K = 0.23$.
Note that considering in the calculations EC instability as well, the conductive EC threshold curve is located at much lower ac voltage than that for the flexodomains.
The EC branch of the SLC in Fig.~\ref{fig:Calculated_FD2}(a) as well as the behavior of the critical wave number and the obliqueness angle in Figs.~\ref{fig:Calculated_FD2}(b)--(c) look very similar to that obtained at lower ac frequency (Fig.~\ref{fig:flexcond_calc}).

It is clearly seen that the voltage combinations belonging to the FDcond--FDdiel transitions are very high and are by far outside the SLC of EC; so normally it should not be observable. However, if EC is inhibited by some reason, the transition between the two types of FD may become accessible.

We  think that actually this happened during the measurement presented in Fig.~\ref{fig:1oo8_5Hz}(b). Though the ac threshold of EC for 1OO8 at $f=5$ Hz is low at $U_{\mathrm{dc}} = 0$, applying a dc bias increases the ac threshold voltage enormously. The larger the dc bias voltage, the more the threshold increases. Finally, due to this effective inhibition of EC, the voltage range of the FDcond--FDdiel transition has been reached. We are convinced that the flexodomains marked FD in Fig.~\ref{fig:1oo8_5Hz}(b) correspond to the conductive type, while those marked with FDSW are a manifestation of flexodomains of the dielectric type. This statement is supported by the similarities of the theoretical and experimental SLCs as well as the qualitative (or even semiquantitative) agreement of the behaviour of the wave numbers at the FD--FDSW (i.e., at the FDcond--FDdiel) transition.

Summarizing, the outcome of the comparison of the experiments and the theory so far is that there is a good agreement along the whole SLC for low $f$, as well as along the dc branch of the SLC at any $f$. Nevertheless, a serious discrepancy is present for the ac branch of the SLC for high $f$. In the following we attempt to tackle this problem.

\subsection{\label{sec:current} Electrical conductivity measurements }

The  threshold voltages of the electric field induced patterns are
governed by,  besides the control parameters (the waveform,
magnitude and frequency of the applied voltage) and the sample
thickness,  a set of temperature dependent material parameters,
which include the dielectric permittivity, the electrical
conductivity and their anisotropies, the elastic constants, the
viscosities and the flexoelectric coefficients
\cite{Kramer1996,Krekhov2008}. Most of these material parameters
depend exclusively on the chemical structure of the compound and
are expected to be identical for various samples kept under the
same conditions. The electrical conductivity, however, originates
in the ionic and electrolytic impurities. Therefore, the actual
value of the conductivity and its anisotropy may vary from sample
to sample depending on the concentration and type of charge
carriers. In addition, the conductivity may change in time under
applied voltage due to, e.g., reversible ionic adsorption at
electrodes, irreversible ionic purification, charge injection
\cite{Bazant2004,Gaspard1973,Murakami1997-1,Murakami1997-2,Huang2012};
thus it may depend on the history of the sample. These
conductivity variations may easily reach an order of magnitude.
Therefore, it is important to obtain information on the
conductivity at the time of pattern formation studies.

In order to conform to this requirement, the electric current flowing
through the sample and the applied voltage signals were recorded simultaneously to image capturing.
From the ac voltage and current components, the complex impedance of the cell (interpreted as a parallel RC circuit) was
determined; it was phase sensitively decomposed into the parallel
resistance $R_{\mathrm{p}}$ and parallel capacitance $C_{\mathrm{p}}$ of the cell. As planar samples were used, for voltages below the pattern onset, one finds
$R_{\mathrm{p}}=(\sigma_{\bot} A/d)^{-1}$  and $C_{p}=\varepsilon_{0}\varepsilon_{\bot} A/d$ (here $A$ is the
effective electrode area); that allows to determine the conductivity $\sigma_{\bot}$ and dielectric permittivity $\varepsilon_{\bot}$.
Above pattern onset, the resistance $R_{\mathrm{p}}$ and the capacitance $C_{\mathrm{p}}$ depend, in addition, on $\sigma_{\mathrm{a}}$, on $\varepsilon_{\mathrm{a}}$, on the director distortion and on the velocity field.

\begin{figure*}[t]
\begin{center}
\includegraphics[width=16cm]{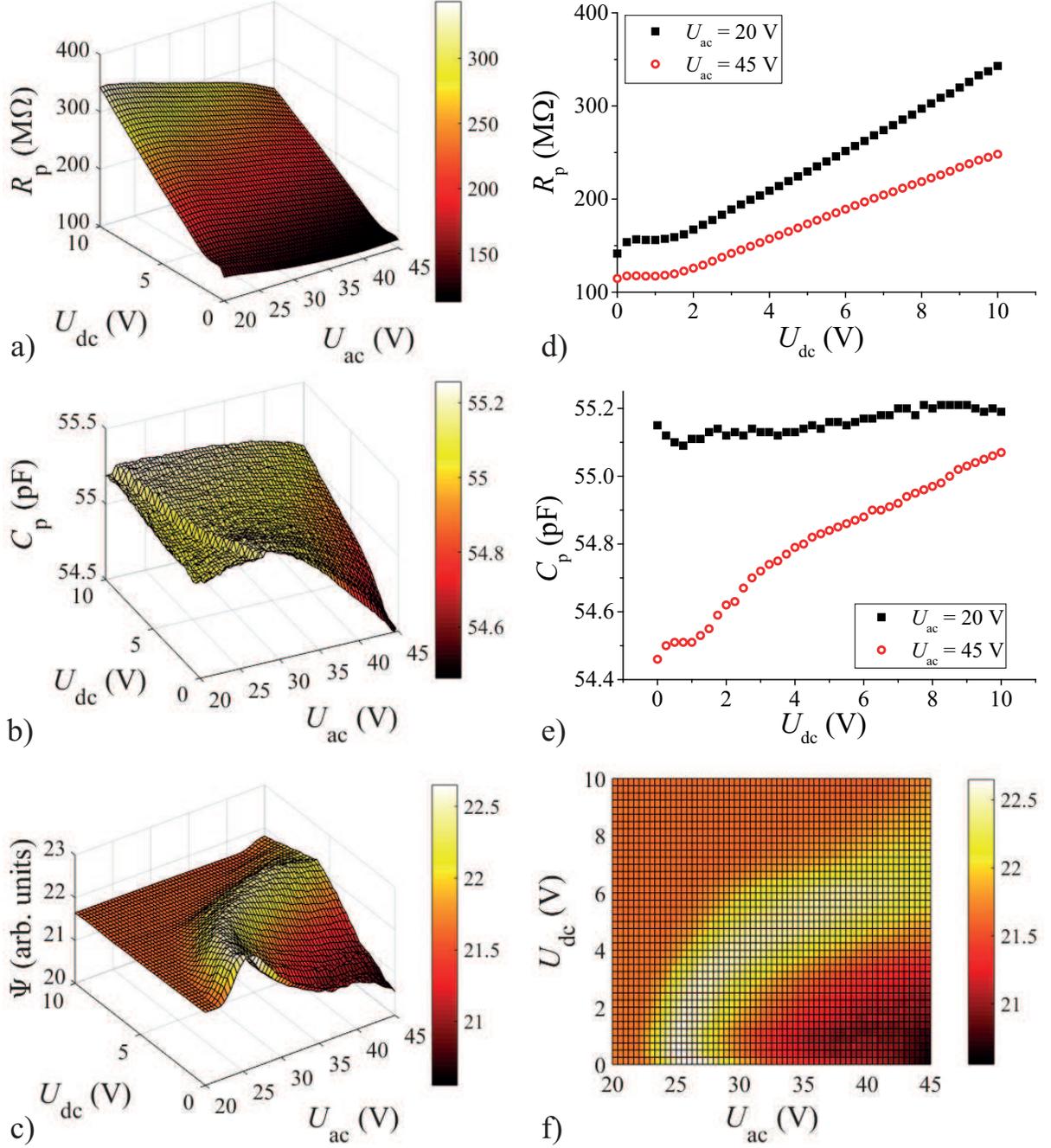}
\end{center}
\caption{(Color online) In the left panel, voltage dependence of a) the parallel
resistance $R_{\mathrm{p}}$, b) the parallel capacitance $C_{\mathrm{p}}$ of a 1OO8 sample at $f=80$ Hz; c) morphological phase diagram (voltage dependence of
the contrast $\Psi$) of the same sample. All 3d graphs are colour coded:
black corresponds to the lowest, white to the highest values. In the right panel, 2d cross-sections of the 3d images (d) for $R_p$, (e) for $C_p$; (f) 2d projection of $\Psi$. The dashed line corresponds to the stability limiting curve of EC. }
\label{fig:1oo8_RpCp_3d}
\end{figure*}

In Fig.~\ref{fig:1oo8_RpCp_3d}, we present the results of the measurements on a $d=19.5$ $\mu$m thick cell.
Figures~\ref{fig:1oo8_RpCp_3d}(a) and \ref{fig:1oo8_RpCp_3d}(b) depict
the ac and dc voltage dependence of $R_{\mathrm{p}}$ and $C_{\mathrm{p}}$, respectively, obtained
during a voltage scan of a ($U_{\mathrm{ac}},U_{\mathrm{dc}}$) area, which includes
the ac branch of the SLC. Figures~\ref{fig:1oo8_RpCp_3d}(d) and \ref{fig:1oo8_RpCp_3d}(e) show a cross-section at fixed $U_{\mathrm{ac}}$. It can be seen that a dc bias voltage
$U_{\mathrm{dc}}\gtrsim 2$ V causes a substantial, nearly linear increase
of $R_{\mathrm{p}}$ (a reduction of $\sigma_{\bot}$), while increasing the ac
voltage results in some decrease of $R_{\mathrm{p}}$ [see also Fig. \ref{fig:1oo8_RpCp_3d}(d)]. On the contrary,
$C_{\mathrm{p}}$ remains nearly constant upon changing $U_{\mathrm{dc}}$; the maximal deviation ($<2$\%) occurs
in the range of small dc, but large ac voltages. This is, however,
the range which is already much above the onset of EC (i.e., it is
a turbulent state) and therefore director distortions are there
non-negligible. This latter statement is supported by the
morphological diagrams in Figs.~\ref{fig:1oo8_RpCp_3d}(c) and \ref{fig:1oo8_RpCp_3d}(f), which depict
the voltage dependence of the pattern contrast $\Psi$ for the same
voltage scan. Some decrease of the contrast at high ac and low dc voltages below the background of the undistorted state is expected to be
due to turbulent light scattering.

As strong dependence of the conductivity $\sigma_{\bot}$ on the dc
bias voltage has been demonstrated in Figs.~\ref{fig:1oo8_RpCp_3d}(a) and \ref{fig:1oo8_RpCp_3d}(d), further
questions arise about the temporal dynamics of the conductivity
change, and whether the anisotropy of the conductivity depends on
the dc voltage too. To answer these questions, high precision
impedance measurements were performed using a dedicated dielectric analyzer.

In a planar cell, $\varepsilon_{\bot}$ and $\sigma_{\bot}$ are the
directly measurable quantities. In order to get information on the
anisotropy, an additional magnetic field can be applied normal to
the electrodes, in order to induce a splay Freedericksz transition
into a quasi-homeotropic state. The threshold magnetic induction
for the cells used in Sec. \ref{sec:acdc} ($d=19.5$ $\mu$m) is
$B_{\mathrm{F}}\approx 0.43$~T at $U_{\mathrm{dc}}=0$; thus the
maximum applicable induction ($B_{\mathrm{max}}=1$ T) is too small
for a reliable estimation of $\varepsilon_{\parallel}$ and
$\sigma_{\parallel}$. For the cell in Sec. \ref{sec:acac} ($d=50$
$\mu$m) one finds $B_{\mathrm{F}}= 0.17$~T at $U_{\mathrm{dc}}=0$;
so the permittivity $\varepsilon(\mathrm{1T})$ measured at the
maximal $B$ is already much closer to $\varepsilon_\parallel$,
yielding $\Delta \varepsilon = \varepsilon
(\mathrm{\mathrm{1T}})-\varepsilon (\mathrm{\mathrm{0T}})\approx
0.90 \, \varepsilon_{\mathrm{a}}$. However, as the dc bias voltage
has a dielectric stabilizing effect
($\varepsilon_{\mathrm{a}}<0$), $B_{\mathrm{F}}$ increases as the
dc bias voltage is increased. Therefore, $\Delta \varepsilon$ and
similarly $\Delta \sigma =
\sigma(\mathrm{\mathrm{1T}})-\sigma(\mathrm{\mathrm{0T}})$ cease
to be good approximations of $\varepsilon_{\mathrm{a}}$ and
$\sigma_{\mathrm{a}}$, respectively, for $U_{dc}\gtrsim 5$ V.

In order to overcome this problem and allow studying the dc  bias
dependence of $\sigma_{\bot}$ and $\sigma_{\parallel}$ in a wide
voltage range ($0<U_{\mathrm{dc}}<40$ V), a cell with $d=1$ mm was
used. At such a thickness, without dc bias $B_{\mathrm{F}}<0.01 \,
B_{\mathrm{max}}$; thus the magnetic field can provide
a uniform bulk orientation $\mathbf{n} \| \mathbf{B}$. Therefore,
rotating the cell in the high magnetic
field alternately to the positions with $\mathbf{B}$ parallel to
and perpendicular to the substrates, $\varepsilon_{\bot}$, $\sigma_{\bot}$ and
$\varepsilon_{\parallel}$, $\sigma_{\parallel}$ could be measured, respectively. The
measurements were performed at $f=1$ kHz with a probing ac rms
voltage of 0.2 V.

\begin{figure}[!h]
\begin{center}
\includegraphics[width=8cm]{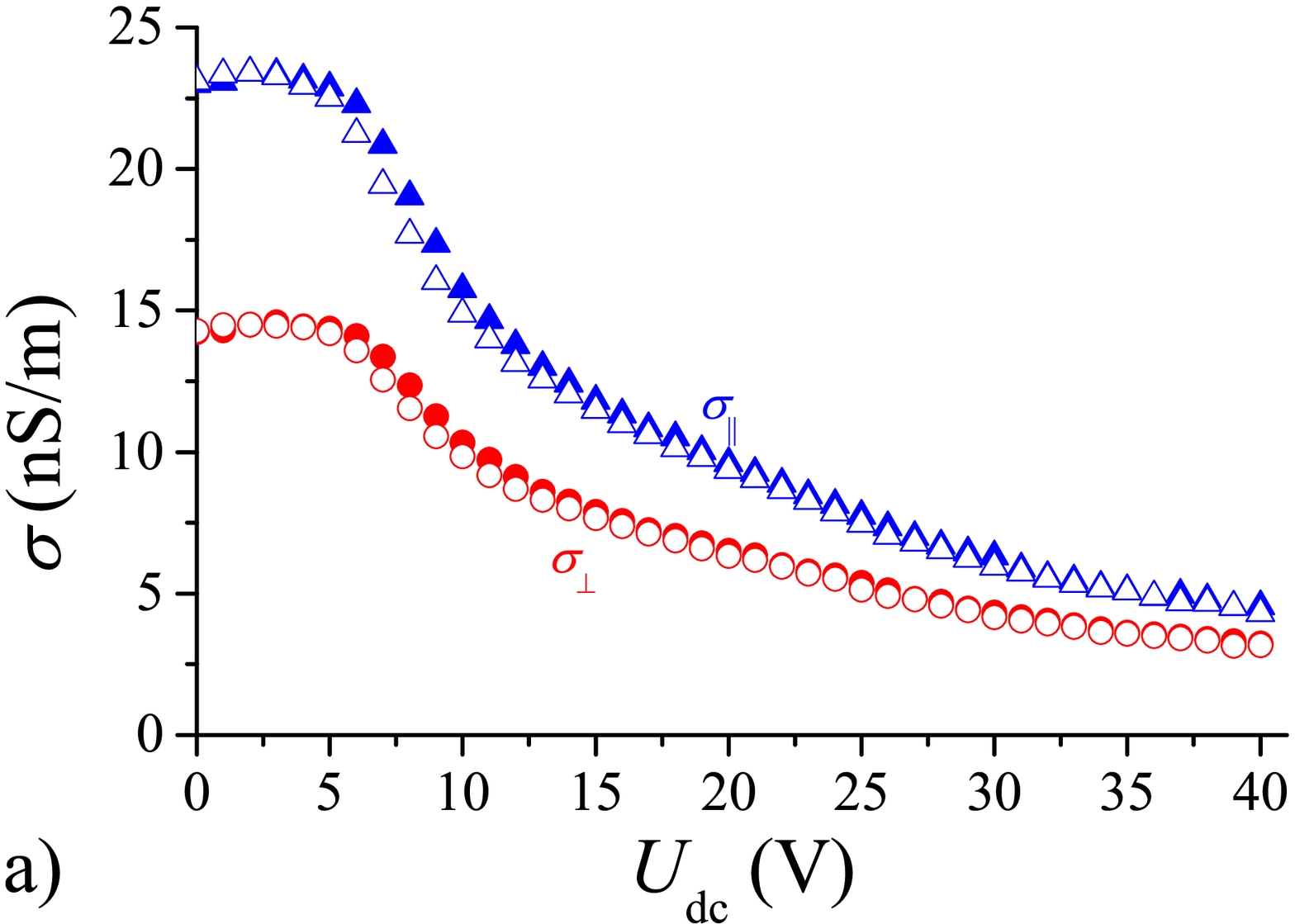} \\  
\includegraphics[width=8cm]{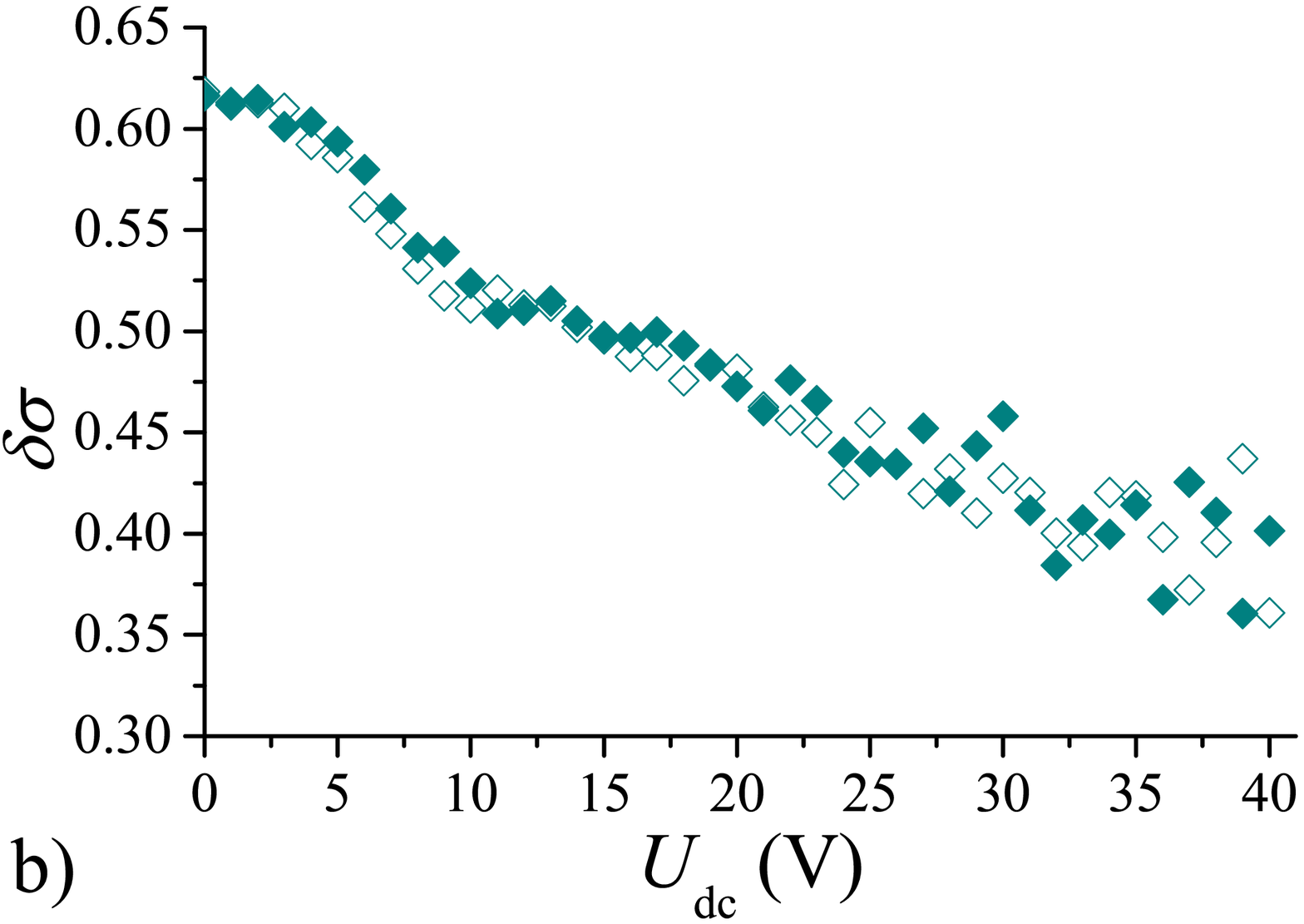}  
\end{center}
\caption{(Color online) Dc voltage dependence of (a) the
conductivities $\sigma_{\bot}$ and $\sigma_{\parallel}$, and (b) the relative conductivity anisotropy $\delta \sigma = (\sigma_{\parallel} - \sigma_{\bot})/\sigma_{\bot}$. The solid (open) symbols plotted at the same $U_{\mathrm{dc}}$ correspond to data measured at the beginning (end) of the interval of about 8 minutes at which the dc voltage was kept constant.}
\label{fig:1oo8_sigma_Vdependence}
\end{figure}

Figures~\ref{fig:1oo8_sigma_Vdependence}(a)--(b) exhibit the $U_{\mathrm{dc}}$
dependence of $\sigma_{\bot}$ and
$\sigma_{\parallel}$, and the relative conductivity anisotropy
$\delta \sigma= \sigma_{\mathrm{a}}/\sigma_{\bot}$, respectively.
The dc voltage was increased gradually in intervals of about 8
minutes; at each voltage the cell was repetitively rotated in the
magnetic field of $B=1$~T in order to obtain the two components of
the conductivity and permittivity. The solid and open symbols
plotted at the same $U_{\mathrm{dc}}$ correspond to data measured
at the beginning and at the end of these intervals, respectively.

It is clear from Fig.~\ref{fig:1oo8_sigma_Vdependence}(a) that
while the conductivity is unaffected by a small dc bias voltage
($U_{\mathrm{dc}}\lesssim 4$ V), it strongly reduces when $U_{\mathrm{dc}}$
increases. The same features could be seen before in Figs.~\ref{fig:1oo8_RpCp_3d}(a) and \ref{fig:1oo8_RpCp_3d}(d), though
there (in a thinner cell) the critical dc bias voltage, which
induces conductivity variations was lower ($U_{\mathrm{dc}}\lesssim 2$ V).
Meanwhile, both permittivity components remained unaffected by the
dc voltage, indicating no dc bias induced change of the
homogeneous planar or quasi-homeotropic director orientation. The
relative conductivity anisotropy $\delta \sigma$ is presented in
Fig.~\ref{fig:1oo8_sigma_Vdependence}(b). It shows that not only
the conductivity components, but $\delta \sigma$ too exhibit a
monotonic decrease upon increasing $U_{\mathrm{dc}}$: at
$U_{\mathrm{dc}}=40$ V, $\delta \sigma$ is just half of its value
at $U_{\mathrm{dc}}=0$.

\begin{figure}[!h]
\begin{center}
\includegraphics[width=8cm]{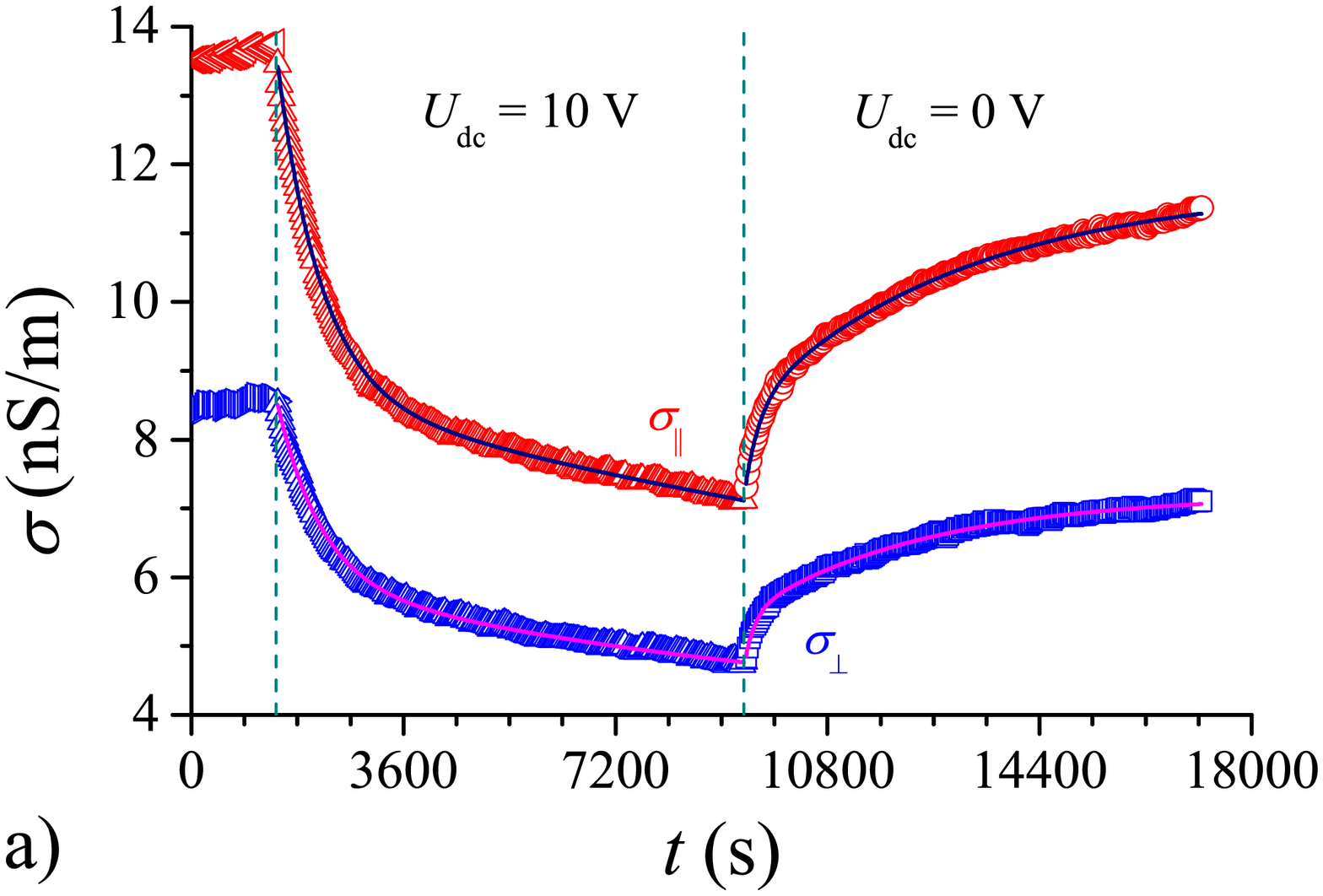} \\  
\includegraphics[width=8cm]{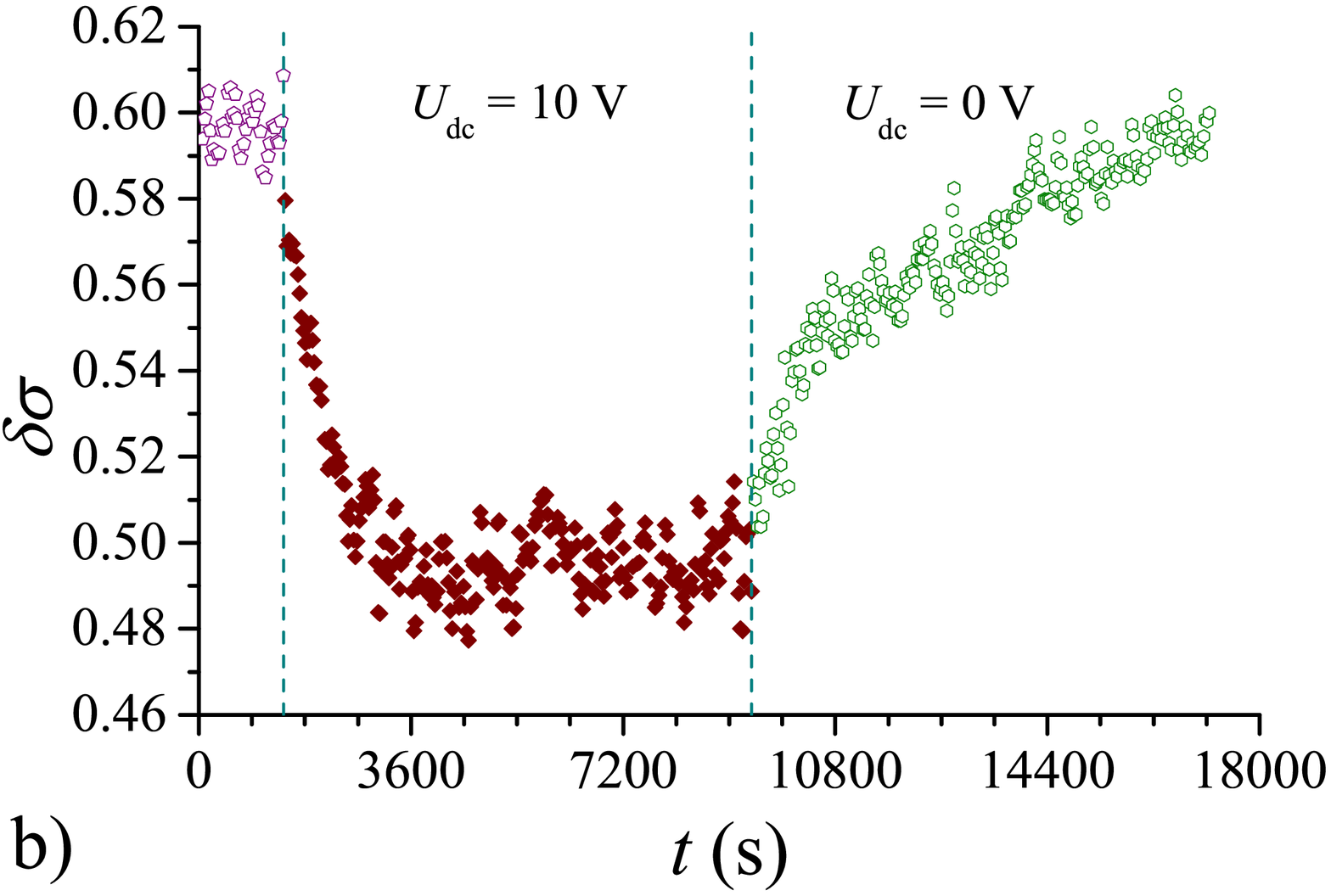}  
\end{center}
\caption{(Color online) Temporal variation of (a) the conductivity
$\sigma_{\bot}$ and $\sigma_{\parallel}$, and (b) the relative
conductivity anisotropy $\delta \sigma =\sigma_{\mathrm{a}}/\sigma_{\bot}$ during step-like increase of the dc voltage
from $U_{dc}=0$ to $U_{dc}=10$ V, and then from $U_{dc}=10$ V to
$U_{dc}=0$. The vertical dashed lines mark the moments of the voltage jumps. The solid lines correspond to fits by a superposition of two exponential decay functions according to Equation \ref{eq:2exp}.} \label{fig:1oo8_sigma_jump}
\end{figure}

Dynamics of the dc voltage induced changes in the conductivity and
in its anisotropy was also studied. Figures~\ref{fig:1oo8_sigma_jump}(a) and \ref{fig:1oo8_sigma_jump}(b)
exhibit the temporal evolution of the conductivities
$\sigma_{\bot}$, $\sigma_{\parallel}$, and the relative
conductivity anisotropy $\delta \sigma$, respectively, following a
dc voltage jump from $U_{\mathrm{dc}}=0$ to $U_{\mathrm{dc}}=10$ V
and vice versa, for the same $d=1$ mm thick sample. It is clearly
seen that neither the change of the conductivity, nor of its
anisotropy is instantaneous; instead, the dc voltage jump
initiates a long lasting relaxation process, which is characterized
by several time scales. The fastest one (not captured by the
figures) is about seconds; the slower ones are from minutes to
hours. For example, the curves in Fig.~\ref{fig:1oo8_sigma_jump}(a) could be well fitted (see the solid
lines) by a superposition of two exponential decay curves:
\begin{equation}
\sigma (t) = \sigma_0 + \sigma_1 \exp[(t_0-t)/\tau_1]
+\sigma_2 \exp[(t_0-t)/\tau_2],
\label{eq:2exp}
\end{equation}
where $t_0$ is the moment of the voltage jump. The time constants
for the upward voltage jump were found as $\tau_1 \sim 730$ s,
$\tau_2 \sim 10000$ s, while for the downward voltage jump $\tau_1
\sim 200$ s, $\tau_2 \sim 3500$ s. The curves clearly show that
after two hours from the voltage jump the conductivity is still
far from saturation and at longer time recordings additional, even
longer time scales may become noticeable. These temporal changes
one has to be aware of, even when the dc voltage dependence of the
conductivity is measured by a voltage scan technique, as in Fig.~\ref{fig:1oo8_sigma_Vdependence}.

We note that the temporal behavior of the conductivity in a
thinner ($d=50$ $\mu$m thick) cell was very similar, except that the
conductivity values were lower.
Monitoring the conductivity variations there, following a dc
voltage jump from $U_{\mathrm{dc}}=0$ to $U_{\mathrm{dc}}=5$ V and
vice versa, $\tau_1 \sim 200$ s, and $\tau_2 \sim 2200$ s were
obtained and for the best fit a third, shorter time scale $\tau_3
\sim 50$ s was also necessary.

The relevant  time scales for the reversible electrical response
when a voltage $U$ is applied or removed are the relaxation time
of the charge carriers $\tau_{\mathrm{q}} = \varepsilon_0
\varepsilon / \sigma$, the transit (or migration) time
$\tau_{\mathrm{ct}} = d^2/(\mu U)$ and the diffusion time
$\tau_{\mathrm{cd}} = d^2/(\mu k_{\mathrm{B}} T/e)$, where $\mu$
is the average mobility of the charge carriers, $k_B$ is the
Boltzmann constant, and $e$ is the electronic charge
\cite{Turnbull1973,Treiber1995,Bazant2004}. For a nematic layer of
thickness $d = 50$~$\mu$m with the dielectric permittivity of
$\varepsilon = 5$, the conductivity of $\sigma = 5$~nS~m$^{-1}$
and typical values of the mobility $\mu \approx (1 - 4) \cdot
10^{-10}$~m$^2$~(V~s)$^{-1}$, one finds $\tau_{\mathrm{q}} =
0.01$~s, $\tau_{\mathrm{ct}} = 6 - 25$~s, and $\tau_{\mathrm{cd}}
= 245 - 980$~s. The charge diffusion time $\tau_{\mathrm{cd}}$
characterizes the process of diffusion of slow charge carriers
from the bulk into the depleted zones due to the electrical double
layers formed at the electrodes at the time scale
$\tau_{\mathrm{q}}$. The same time $\tau_{\mathrm{cd}}$ is typical
for the equilibration of the charge distribution after the applied
voltage is switched off. Clearly, $\tau_{\mathrm{q}}$ is too small
to be detectable with our experimental technique. The other two
time scales, $\tau_{\mathrm{cd}}$ and $\tau_{\mathrm{ct}}$, are
however, in the order of the experimentally found characteristic
times $\tau_1$ and $\tau_3$.

Other processes, which are much longer in time, are the specific ionic adsorption and diffusion into the polyimide orienting layers covering the cell electrodes \cite{Huang2012,Murakami1997-1,Murakami1997-2}.
Apart from the characteristic times of the reversible dynamics mentioned above, there are certainly some irreversible processes that can also be responsible for the long time conductivity variations shown in Fig.~\ref{fig:1oo8_sigma_jump}.
These are ionic purification of the sample as well as possible decomposition of the nematic molecules and/or dissolution of impurity ions from the polyimide layers \cite{Gaspard1973,Murakami1997-2}.

The characteristic times given above are much longer than the
typical acceptable waiting time between voltage steps in voltage
scans. It means that scanning the $U_{\mathrm{ac}}$--$U_{\mathrm{dc}}$ plane
practically cannot occur under equivalent conditions for the
conductivity, if dc voltage has ever been applied. As a
consequence, $\sigma$ measured at a particular ($U_{\mathrm{ac}}$,
$U_{\mathrm{dc}}$) combination may depend on the route of reaching that
point as well as on the sample history (how large dc bias had been
applied, when and for how long time). This means that exploring
the $U_{\mathrm{ac}}$--$U_{\mathrm{dc}}$ plane along horizontal lines (increasing
$U_{\mathrm{ac}}$ at constant $U_{\mathrm{dc}}$) or along vertical lines (increasing
$U_{\mathrm{dc}}$ at constant $U_{\mathrm{ac}}$) might not be equivalent.

\section{\label{sec:discussion} Discussion}

The increase of the ac thresholds of EC patterns upon applying dc
bias voltage seen in Figs. \ref{fig:1oo8_5Hz} and
\ref{fig:1oo8_RpCp_3d}(f) clearly does not match the theoretical
predictions in Figs. \ref{fig:flexcond_calc}(a), \ref{fig:flexdiel_calc}(a) and
\ref{fig:Calculated_FD2}(a). The latter were calculated, however,
in the framework of a model where $\sigma_{\bot}$ and $\sigma_{\mathrm{a}}/\sigma_{\bot}$ are independent of $U_{\mathrm{dc}}$, while the experiments have proved that the conductivity and its anisotropy change substantially if a dc bias is applied.

Although a dc-bias dependent conductivity is not captured by our extended SM model based on the Maxwell equations in the quasi-static approximation assuming ohmic conductivity of the nematic, one can verify the influence of various conductivity and/or conductivity anisotropy on the critical voltages of EC patterns.
Note that the theoretical thresholds of flexodomains are independent of $\sigma_{\bot}$ and $\sigma_{\mathrm{a}}/\sigma_{\bot}$, since for FD pattern one has $\mathbf{q}_{\mathrm{FD}} \bot \mathbf{n}_0$ resulting in $\varrho_{\mathrm{e}} = 0$.

\begin{figure}[!h]
\begin{center}
\includegraphics[width=8cm]{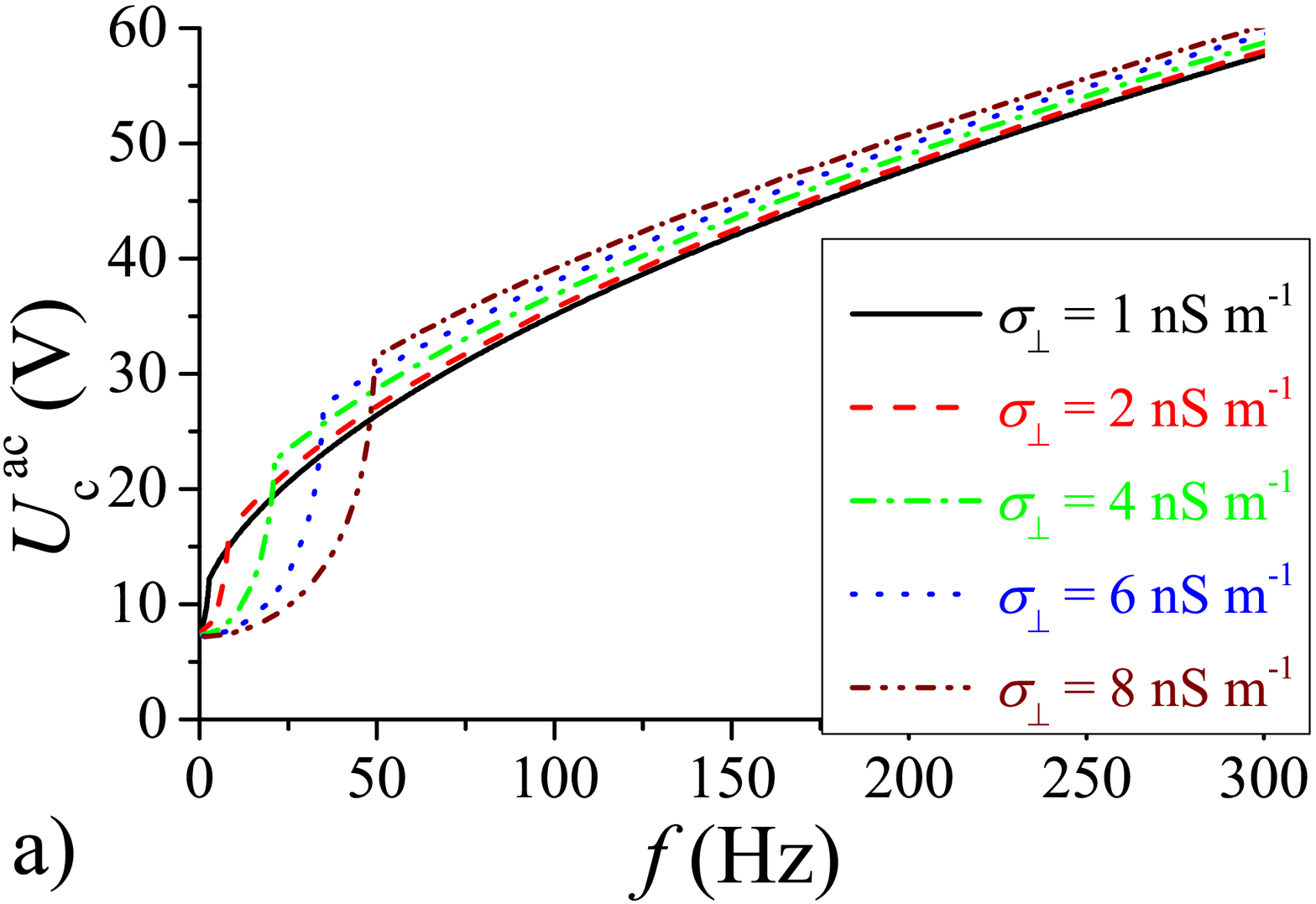}  
\includegraphics[width=8cm]{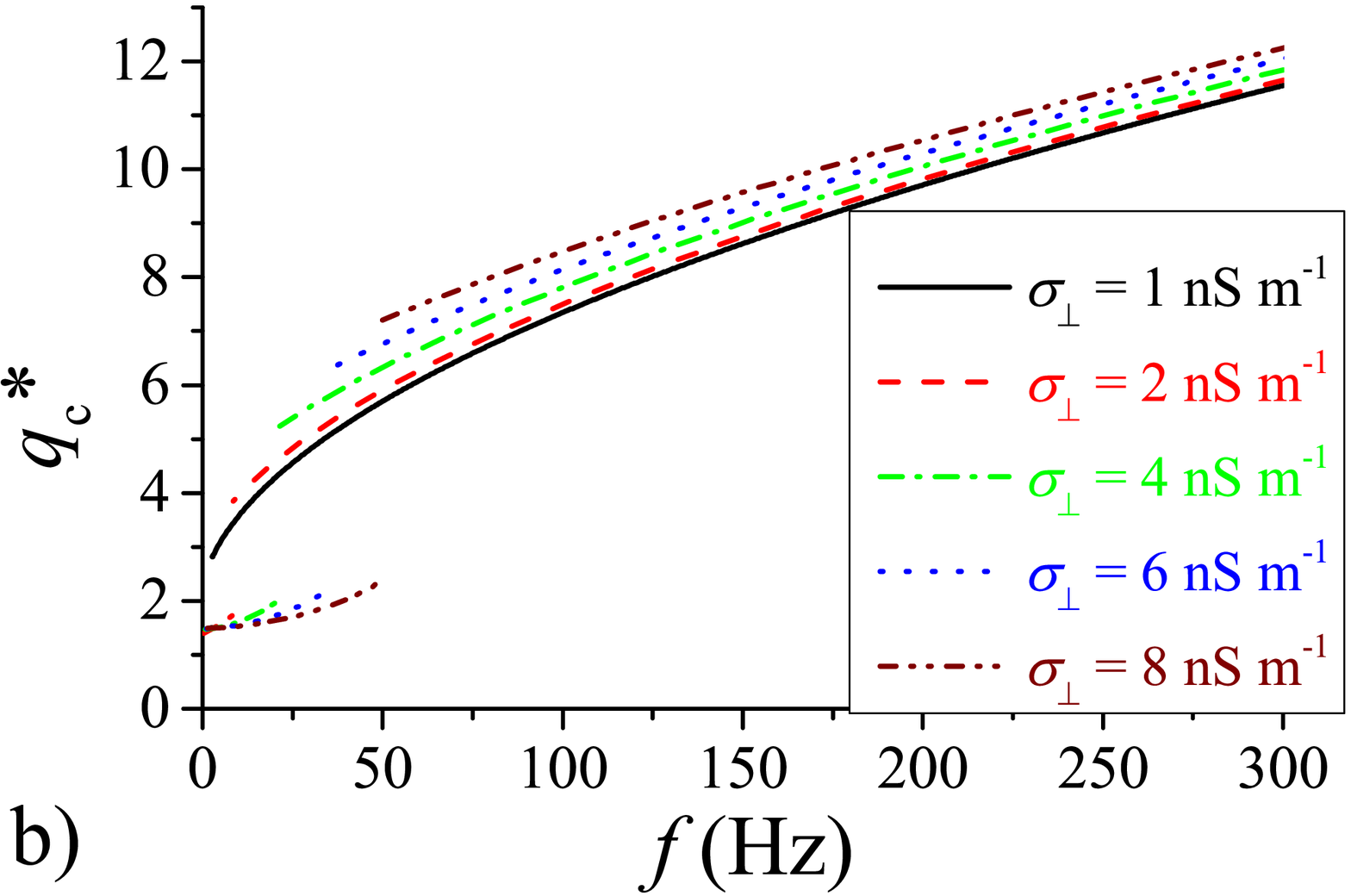}  
\includegraphics[width=8cm]{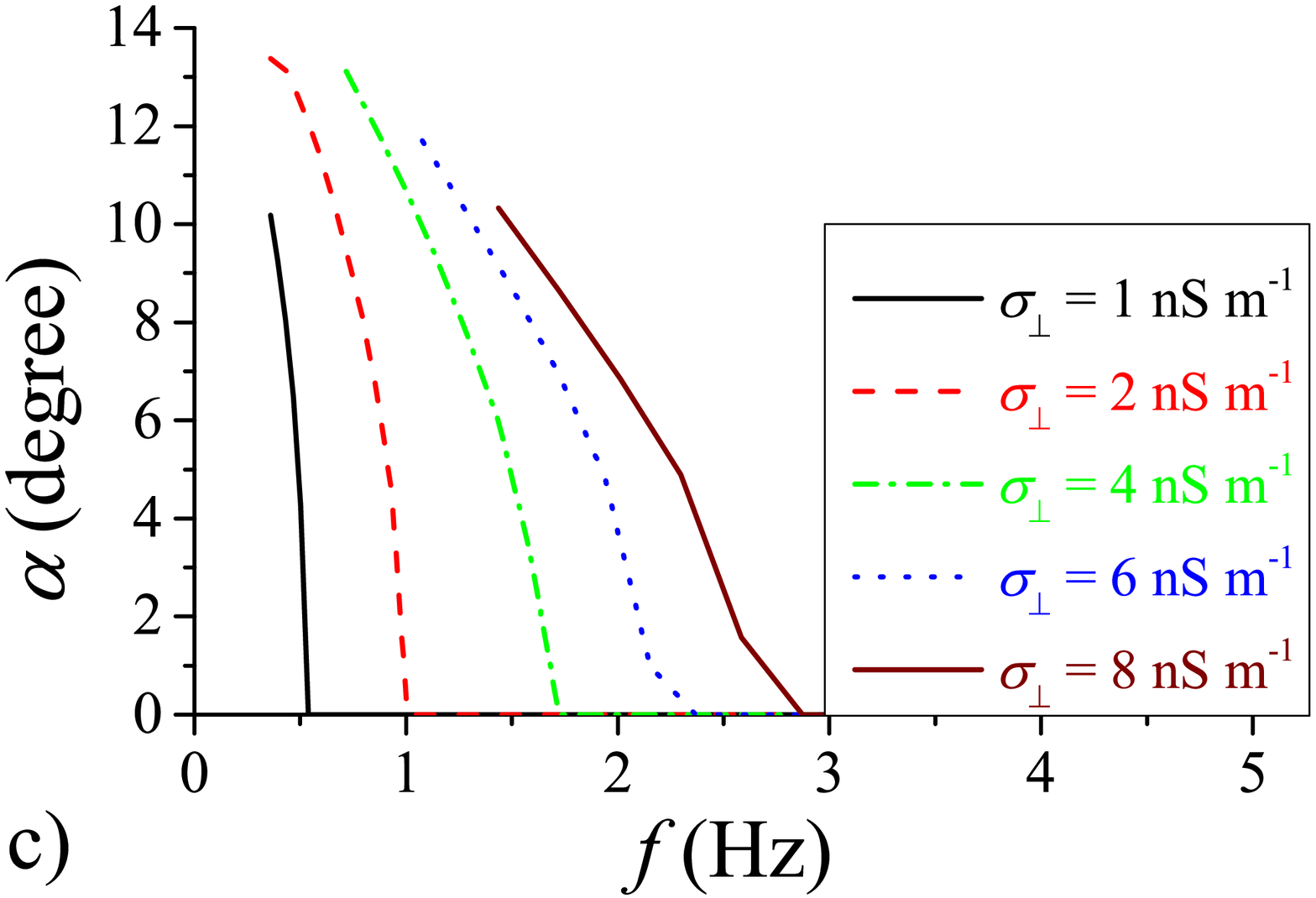}  
\end{center}
\caption{(Color online) Calculated frequency dependence of a) the threshold voltages $U_{\mathrm{c}}^{\mathrm{ac}}$, b) the dimensionless critical wave number $q_{\mathrm{c}}^*$ and c) the obliqueness angle $\alpha$ of EC patterns at pure ac driving at the conductivity anisotropy of $\sigma_{\mathrm{a}}/\sigma_{\bot} = 0.5$ and different values of the conductivity $\sigma_{\bot}$.}
\label{fig:Uac(f)sigma}
\end{figure}

\begin{figure}[!h]
\begin{center}
\includegraphics[width=8cm]{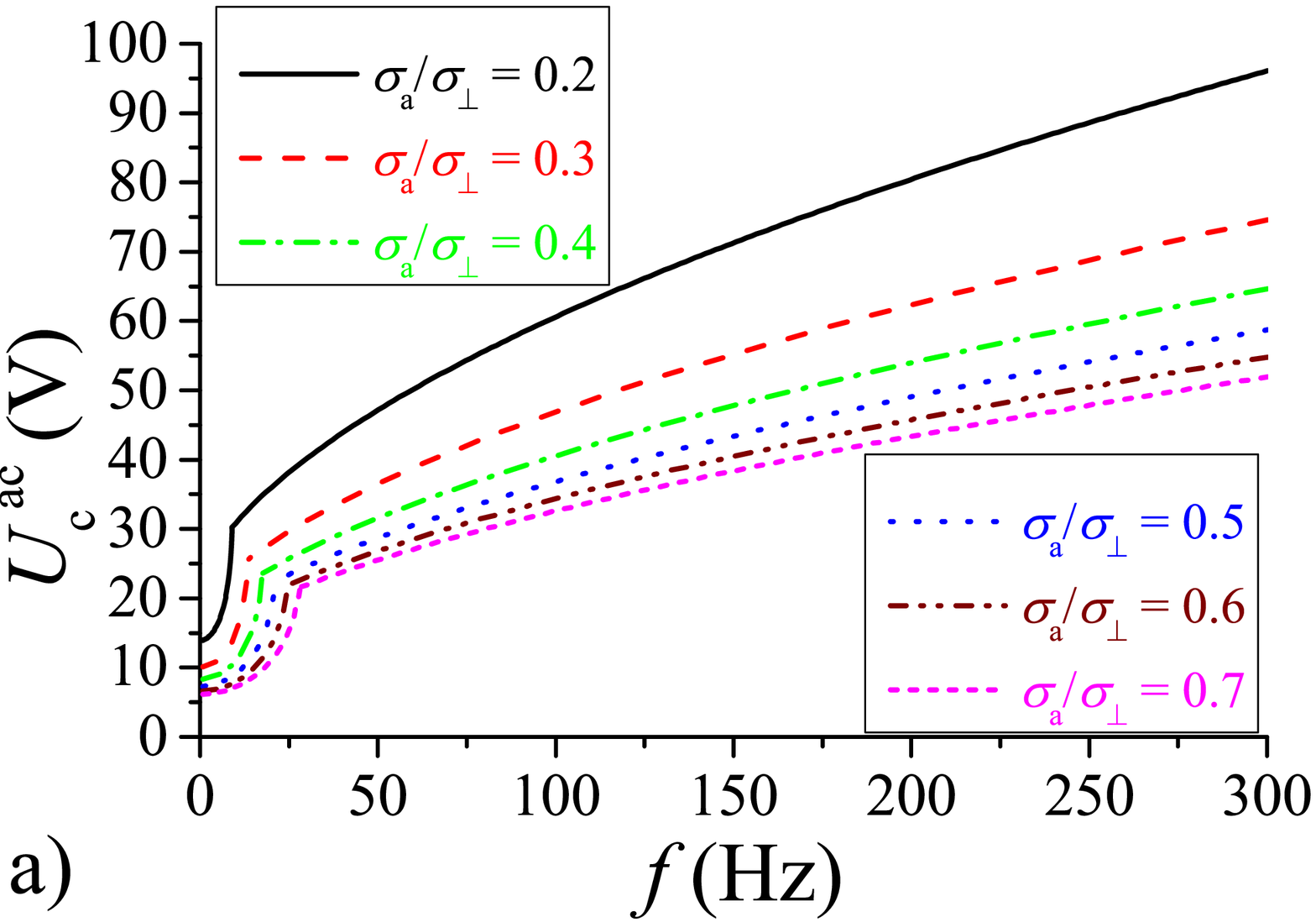}  
\includegraphics[width=8cm]{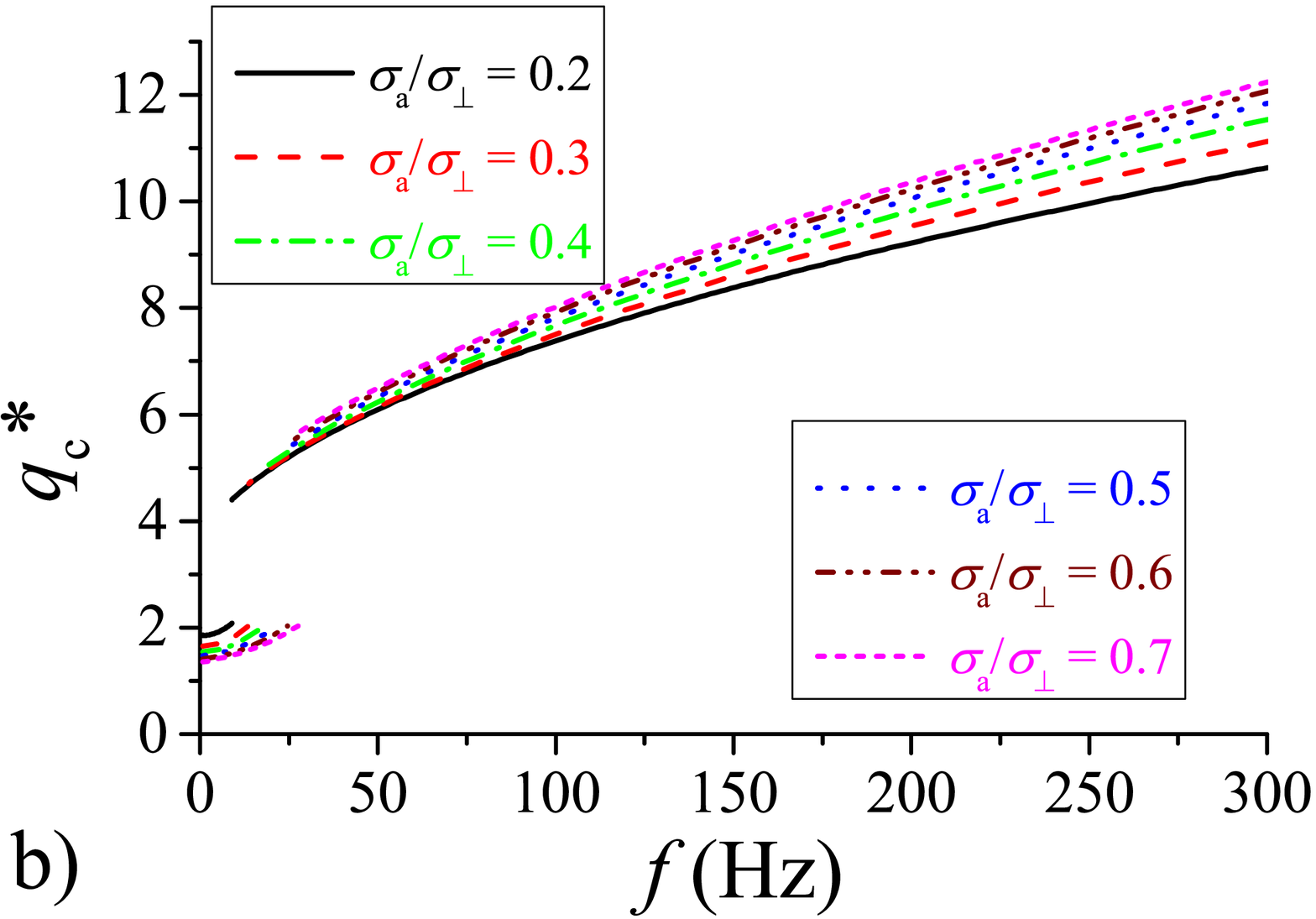}  
\includegraphics[width=8cm]{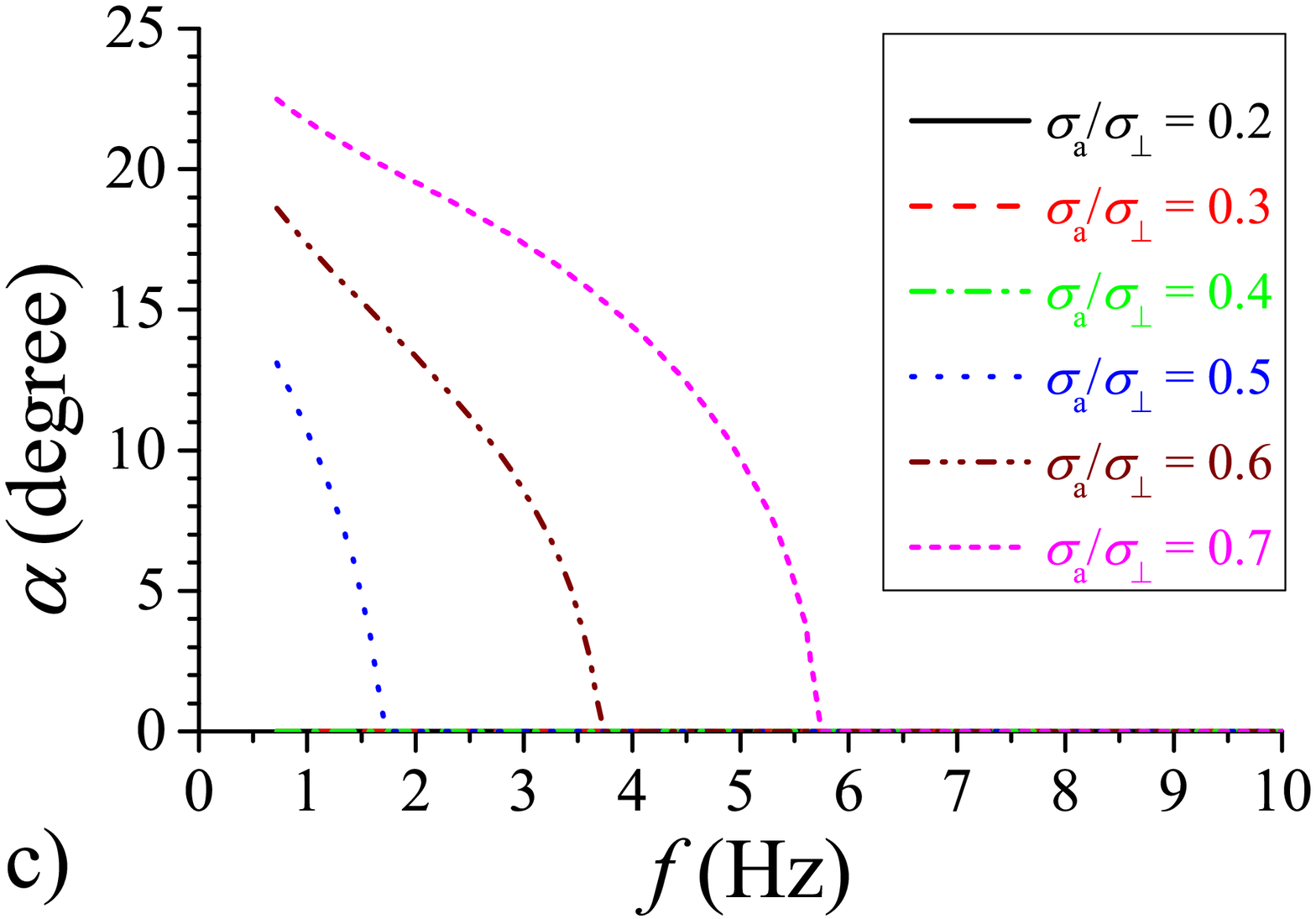}  
\end{center}
\caption{(Color online) Calculated frequency dependence of a) the threshold voltages $U_{\mathrm{c}}^{\mathrm{ac}}$, b) the dimensionless critical wave number $q_{\mathrm{c}}^*$ and c) the obliqueness angle $\alpha$ of EC patterns at pure ac driving at the conductivity of $\sigma_{\bot} = 4$ nS~m$^{-1}$ and different values of the conductivity anisotropy $\sigma_{\mathrm{a}}/\sigma_{\bot}$.}
\label{fig:Uac(f)anisotropy}
\end{figure}

Figures~\ref{fig:Uac(f)sigma}(a)--\ref{fig:Uac(f)sigma}(c) show the frequency dependence of the threshold ac voltages $U_{\mathrm{c}}^{\mathrm{ac}}$, the dimensionless critical wave number $q_{\mathrm{c}}^*$ and the obliqueness angle $\alpha$, respectively, for the
conductive and dielectric regimes of EC at pure ac driving,
calculated for the material parameter set of
1OO8 given in Section \ref{sec:anticipation}, except $\sigma_{\mathrm{a}}/\sigma_{\bot} = 0.5$ and for different $\sigma_{\bot}$ values taken from within the range found for thin samples. Similarly, Figs.~\ref{fig:Uac(f)anisotropy}(a)--\ref{fig:Uac(f)anisotropy}(c) show the frequency dependence of $U_{\mathrm{c}}^{\mathrm{ac}}$, $q_{\mathrm{c}}^*$ and $\alpha$, respectively, however, for $\sigma_{\bot}=4$ nS/m and various $\sigma_{\mathrm{a}}/\sigma_{\bot} $ values.

First, in Figs.~\ref{fig:Uac(f)sigma}(a) and \ref{fig:Uac(f)anisotropy}(a), one can notice that the
crossover frequency $f_{\mathrm{c}}$ of the transition from conductive EC to dielectric EC shifts toward lower $f$ either if
$\sigma_{\bot}$ or $\sigma_{\mathrm{a}}/\sigma_{\bot} $ is diminished. This tendency may explain the
experimentally found behaviour shown in Fig. \ref{fig:1oo8_Vac(f,Vdc)}:
the dc bias yields the decrease of $f_{\mathrm{c}}$ and finally the
disappearance of the conductive regime via reducing the
conductivity and/or its anisotropy. The dc-bias-induced conductive to dielectric
transition observed at $f=10$ Hz [EC stability branch in Fig. \ref{fig:1oo8_5Hz}(c)] is
another manifestation of the same mechanism: the crossover frequency shifts from
above to below the frequency of the ac voltage component due to the change in
the conductivities.

From Figs.~\ref{fig:Uac(f)sigma}(c) and \ref{fig:Uac(f)anisotropy}(c), it follows that the Lifshitz--point $f_{\mathrm{L}}$ (the frequency of the oblique-to-normal roll transition, where $\alpha$ becomes 0) behaves in a similar way: $f_{\mathrm{L}}$ shifts to lower values if either $\sigma_{\bot}$ or $\sigma_{\mathrm{a}}/\sigma_{\bot}$ diminishes.

Let us now focus on the threshold voltages and the wave numbers. It can immediately be
perceived from Figs.~\ref{fig:Uac(f)sigma}(a) and \ref{fig:Uac(f)sigma}(b) that the two EC regimes
behave differently when changing $\sigma_{\bot}$. In the conductive regime, the reduction of
$\sigma_{\bot}$ leads to a substantial increase of $U_{\mathrm{c}}^{\mathrm{ac}}$, while in the dielectric one, it results in a minor reduction of the threshold.
The critical wave number follows the same trend as the threshold: higher $U_{\mathrm{c}}^{\mathrm{ac}}$ is accompanied by higher $q_{\mathrm{c}}^*$.
In contrast to this, Fig.~\ref{fig:Uac(f)anisotropy}(a) shows that a reduction of $\sigma_{\mathrm{a}}/\sigma_{\bot}$ yields an increase of $U_{\mathrm{c}}^{\mathrm{ac}}$ both in the conductive and the dielectric EC regimes. According to Fig.~\ref{fig:Uac(f)anisotropy}(b), $q_{\mathrm{c}}^*$ follows the trend of $U_{\mathrm{c}}^{\mathrm{ac}}$ only in the conductive regime; in the dielectric one smaller $\sigma_{\mathrm{a}}/\sigma_{\bot}$ results in smaller  $q_{\mathrm{c}}^*$, in spite of the higher $U_{\mathrm{c}}^{\mathrm{ac}}$.

\begin{figure}[!h]
\begin{center}
\includegraphics[width=8cm]{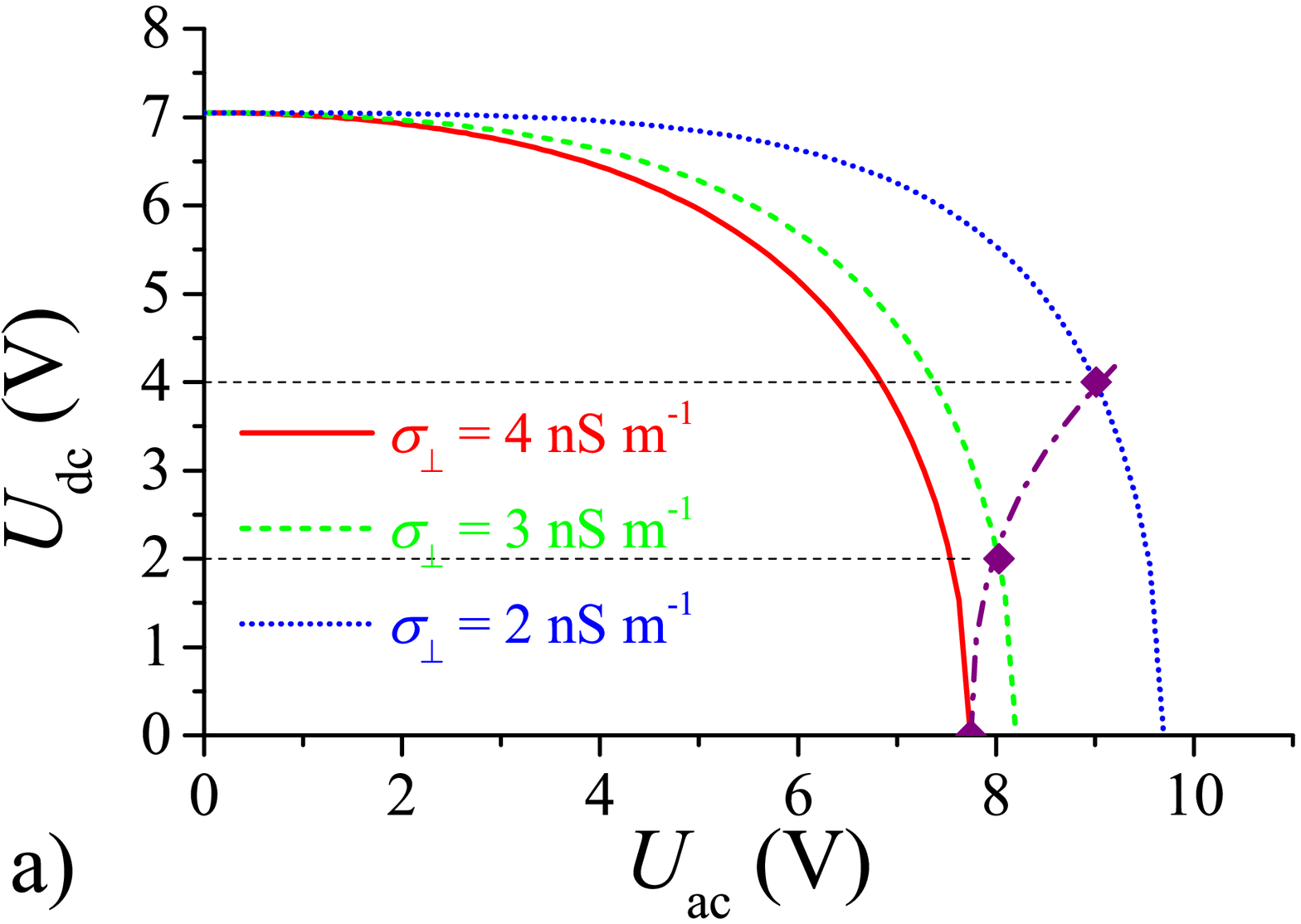}\\
\includegraphics[width=8cm]{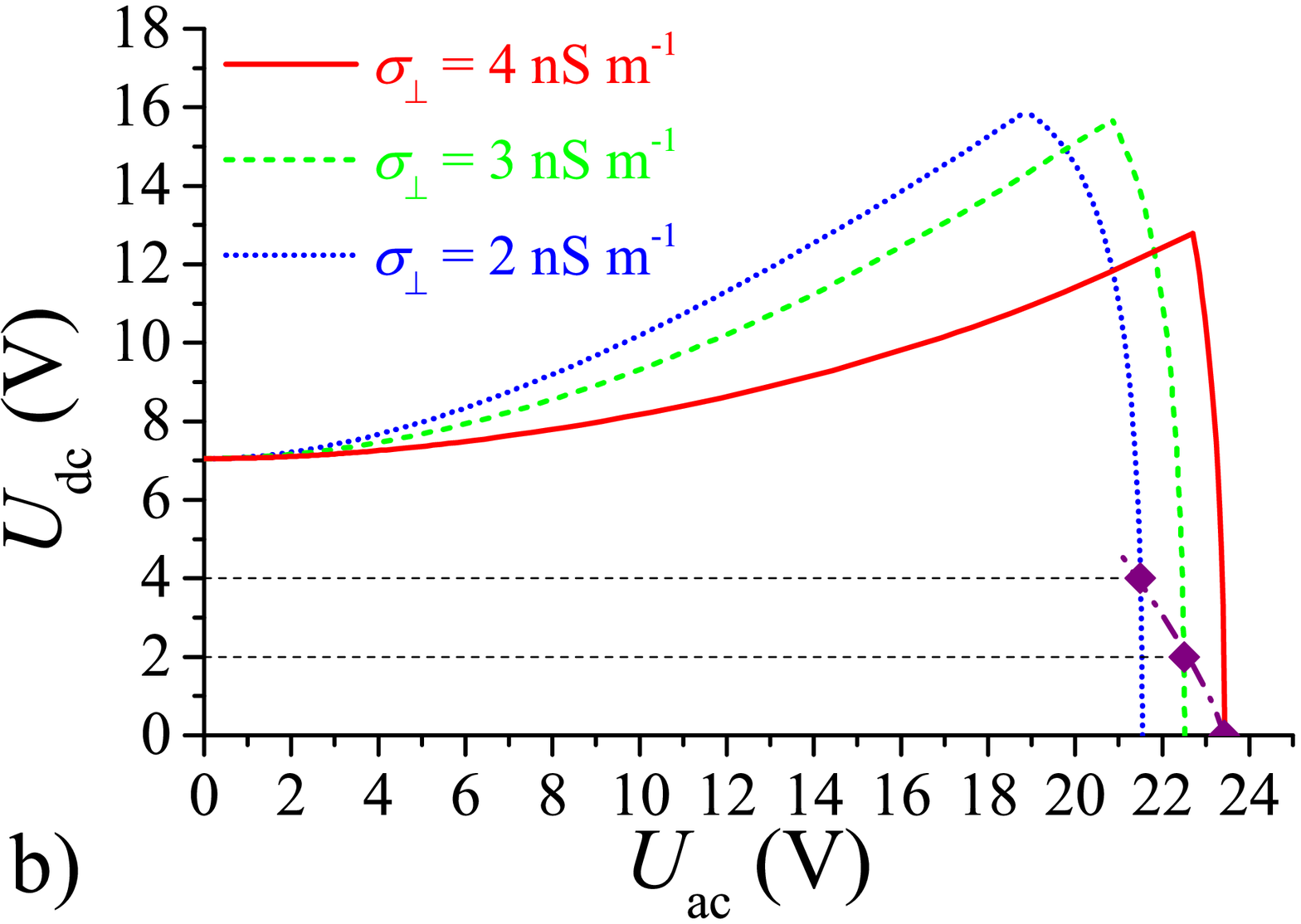}
\end{center}
\caption{(Color online) The stability limiting curves under combined dc and ac voltages in the $U_{\mathrm{ac}}$--$U_{\mathrm{dc}}$ plane for $\sigma_{\mathrm{a}}/\sigma_{\bot}=0.5$  and different conductivity
values $\sigma_{\bot}$, calculated with 1OO8 parameter set for a) the conductive regime ($f=5$~Hz), and b) the
dielectric regime ($f=25$~Hz). Diamonds were the thresholds, if one assumes $\sigma_{\bot}=4$ nS m$^{-1}$ at $U_{\mathrm{dc}}=0$, $\sigma_{\bot}=3$~nS~m$^{-1}$ at $U_{\mathrm{dc}}=2$~V and $\sigma_{\bot}=2$~nS~m$^{-1}$ at $U_{\mathrm{dc}}=4$~V.
The dash-dotted line indicates the trend of the resulting stability limiting curve. } \label{fig:sigma_calc}
\end{figure}

Knowing the behaviour without dc bias, one can also calculate the morphological phase diagram in the $U_{\mathrm{ac}}$--$U_{\mathrm{dc}}$ plane for various
$\sigma_{\bot}$ and $\sigma_{\mathrm{a}}/\sigma_{\bot}$ combinations.
Figures~\ref{fig:sigma_calc}(a) and \ref{fig:sigma_calc}(b) show, how the SLC is affected by $\sigma_{\bot}$, for low $f$ (conductive EC) and high $f$ (dielectric EC), respectively.

\begin{figure}[!h]
\begin{center}
\includegraphics[width=8cm]{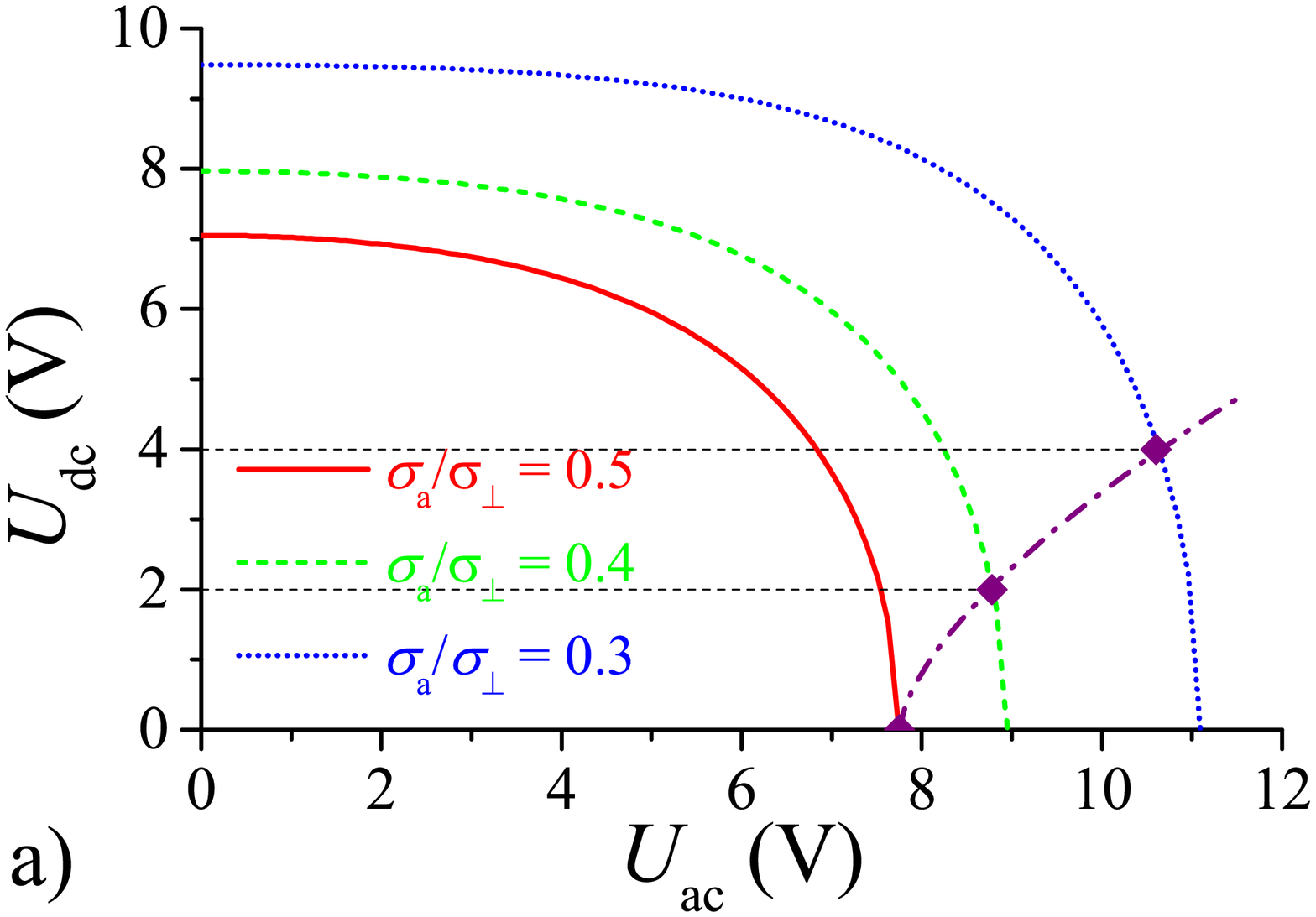}\\
\includegraphics[width=8cm]{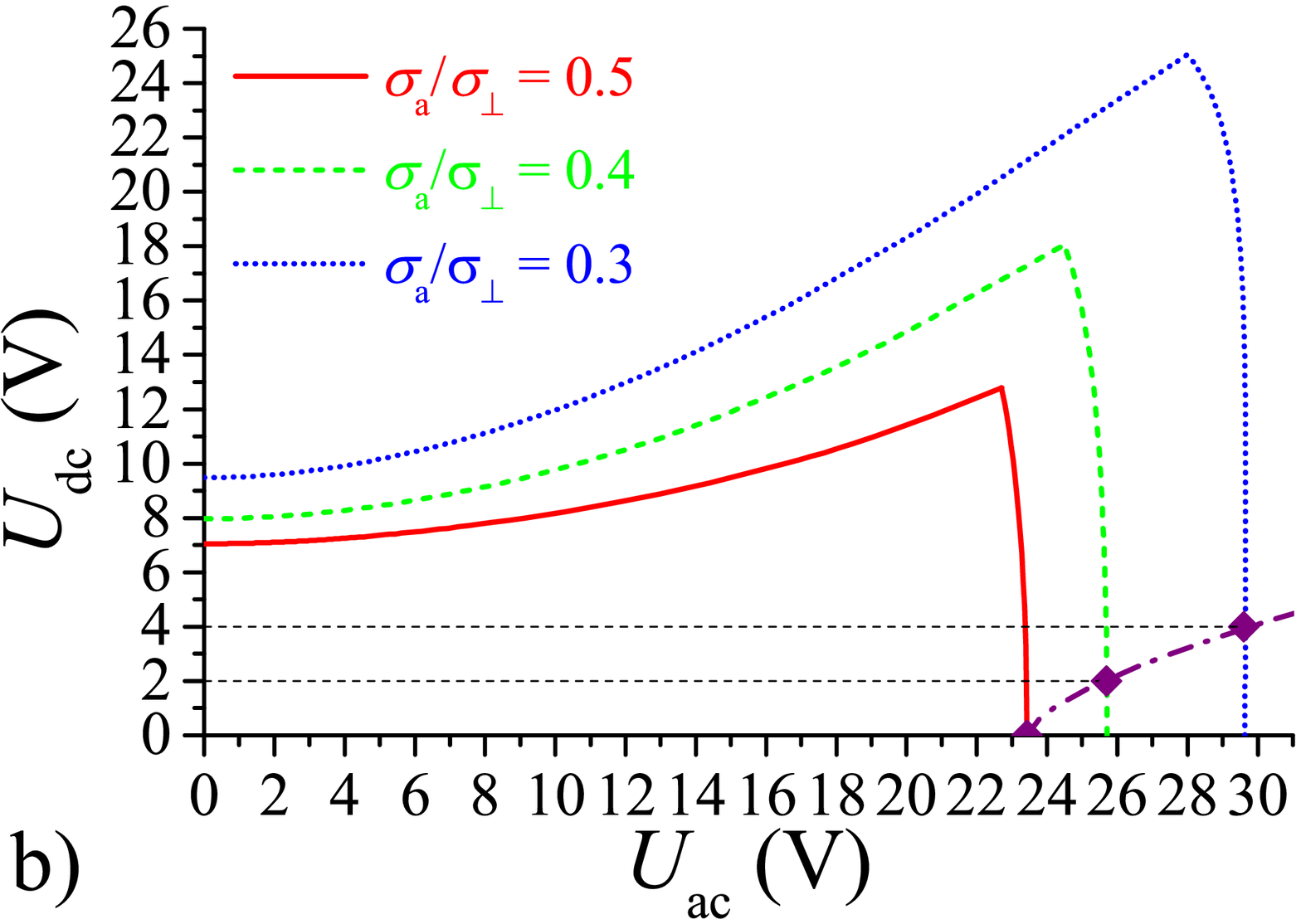}
\end{center}
\caption{(Color online) The stability limiting curves under combined dc and ac voltages in the $U_{\mathrm{ac}}$--$U_{\mathrm{dc}}$ plane for $\sigma_{\bot}=4$~nS~m$^{-1}$ and different conductivity anisotropy
values $\sigma_{\mathrm{a}}/\sigma_{\bot}$, calculated with 1OO8 parameter set for a) the conductive regime ($f=5$ Hz), and b) the
dielectric regime ($f=25$ Hz). Diamonds were the thresholds, if one assumes $\sigma_{\mathrm{a}}/\sigma_{\bot}=0.5$ at $U_{\mathrm{dc}}=0$, $\sigma_{\mathrm{a}}/\sigma_{\bot}=0.4$ at $U_{\mathrm{dc}}=2$ V and $\sigma_{\mathrm{a}}/\sigma_{\bot}=0.3$ at $U_{\mathrm{dc}}=4$ V.
The dash-dotted line indicates the trend of the resulting stability limiting curve.} \label{fig:anisotropy_calc}
\end{figure}

With Figs.~\ref{fig:1oo8_RpCp_3d}(d) and \ref{fig:1oo8_sigma_Vdependence}(a) we have clearly proved that applying a dc bias voltage results in a reduction of $\sigma_{\bot}$. Therefore, if experimental data are going to be compared with the theoretical predictions, for each $U_{\mathrm{dc}}$ the theoretical $U_{\mathrm{ac}}$, $U_{\mathrm{dc}}$ combinations should be taken from the corresponding (and thus different) stability limit curve, as shown by the diamond symbols in Fig.~\ref{fig:sigma_calc}.

In the conductive regime, the SLC shifts to the right when $\sigma_{\bot}$ becomes smaller. This results in an increase of the ac threshold voltage due to the dc-bias-induced conductivity reduction, which may overcome the small threshold decrease expected otherwise at fixed conductivity value [the dash-dotted line in Fig.~\ref{fig:sigma_calc}(a) declines toward right]. This may qualitatively explain the experimentally found [see Fig.~\ref{fig:1oo8_5Hz}(a)] inhibition of the EC pattern formation upon applying dc bias voltage.

We note here that, as seen in Fig.~\ref{fig:1oo8_RpCp_3d}(d), the dc-bias-induced conductivity reduction may be negligible for low dc bias voltage of $U_{\mathrm{dc}}\lesssim 1$--2 V. In such case the above mechanism does not play a role and so the threshold reduction predicted for constant $\sigma_{\bot}$ should become effective. Indeed, it is seen in Fig.~\ref{fig:1oo8_RpCp_3d}(f) that the EC branch of the SLC first declines toward left (smaller $U_{\mathrm{c}}^{\mathrm{ac}}$) and turns back toward right (higher $U_{\mathrm{c}}^{\mathrm{ac}}$) only for $U_{\mathrm{dc}}\gtrsim 1$ V.

In contrast to the conductive one, in the dielectric regime the
reduction of $\sigma_{\bot}$ leads to the decrease of $U_{\mathrm{c}}^{\mathrm{ac}}$.
Therefore the SLCs in Fig.~\ref{fig:sigma_calc}(b) shift towards left upon decreasing $\sigma_{\bot}$. As a consequence, the procedure described above
would reduce $U_{\mathrm{c}}^{\mathrm{ac}}$ even more than expected for a constant
$\sigma_{\bot}$ [the dash-dotted line in Fig.~\ref{fig:sigma_calc}(b) declines toward left].

A similar analyis can be done exploring the effect of the conductivity anisotropies.
Figures~\ref{fig:anisotropy_calc}(a) and \ref{fig:anisotropy_calc}(b) show, how the SLC is affected by $\sigma_{\mathrm{a}}/\sigma_{\bot}$, for low $f$ (conductive EC) and high $f$ (dielectric EC), respectively.
It is seen that the trends are here similar both in the conductive and in the dielectric regimes: the SLC shifts upward and to the right upon the reduction of $\sigma_{\mathrm{a}}/\sigma_{\bot}$ . This means that in the conductive regime the reduction of either $\sigma_{\bot}$ or $\sigma_{\mathrm{a}}/\sigma_{\bot}$ have similar consequences. In contrast, for the frequency of the ac voltage component in the dielectric regime, the SLC shifts to the left upon decreasing $\sigma_\bot$ but to the right when reducing $\sigma_{\mathrm{a}}/\sigma_\perp$.

The impedance measurements have shown that, besides the conductivity value
$\sigma_{\bot}$, the relative conductivity anisotropy is also
affected by the dc bias; $\sigma_{\mathrm{a}}/\sigma_{\bot}$ also reduces with increasing $U_{\mathrm{dc}}$. It is seen in Figs.~\ref{fig:anisotropy_calc}(a) and \ref{fig:anisotropy_calc}(b) that the diamonds corresponding to the thresholds taken from subsequent SLCs (corresponding to the diminishing relative conductivity anisotropy) indicate a net increase of $U_{\mathrm{c}}^{\mathrm{ac}}$ upon increasing the dc bias voltage, for both the conductive and the dielectric regimes (the dash-dotted lines decline toward right for both cases), in agreement with the experimental curves in Fig.~\ref{fig:1oo8_5Hz}.

\section{\label{sec:sum} \bf{Summary}}

In this paper we have reported about the effect of combined
(ac+dc) driving on the electric field induced patterns in a
nematic liquid crystal. We have found that the superposition of ac
and dc voltages typically hinders the pattern forming mechanisms
as it leads to an increase of the critical voltages. Consequently, the
pattern-free region extends to voltages much above the pure ac or
dc thresholds and closure of the stable region could only be observed at
low frequencies. At high ac and dc voltages, a transition between two types of flexodomains could be observed, which is manifested in a sudden change in the wave number of the pattern. This transition could be identified as a crossover between flexoelectric domains of conductive and dielectric time symmetries.

We have found that under combined dc and ac driving, the increase of the critical dc voltage for flexodomains at increasing ac voltage component is well captured by the linear stability analysis of the extended standard model.
In contrast, the increase of the ac threshold for EC patterns by a dc voltage bias is in contradiction with the theoretical predictions.

It is known from studies on isotropic weak electrolytes that
application of a dc voltage may lead to charge depletion and to
building up an electric double layer at the electrodes, which
results in an inhomogeneous electric field with large gradients
across the sample and a nonlinear current response
\cite{Pikinbook,Barbero2006,Derfel2009}. Nematics can also be
regarded as (anisotropic) weak electrolytes, as their small
electrical conductivity is originating from ionic contaminants. Thus
the above effects may occur in nematics too. An experimental proof
of their presence is the measured reduction of the sample's
conductivity upon application of dc bias voltage. The experiments proved that the relative anisotropy of the conductivity is also affected by the dc voltage.

In a lack of theoretical models for weak electrolytes capturing the reduction of the conductivity under application of dc bias voltage, we have analyzed the qualitative trends for the instability thresholds in the framework of the extended SM by varying the model parameters.
For fixed values of the conductivity and the anisotropy of conductivity, the discrepancy between calculated EC instability curves and experiments is not surprising.
We have shown, however, that when taking into account the experimentally found variations of the conductivity and its anisotropy due to the dc bias voltage, all peculiarities of the stability limiting curves for the EC instability can be qualitatively well reproduced.

\section*{\bf{ACKNOWLEDGEMENTS}}
Financial support by the Hungarian Scientific Research Fund (OTKA) grant No. NN110672 is gratefully acknowledged. We wish to thank M. Khazimullin for stimulating discussions and constructive comments on the manuscript, and V.  Kenderesi for technical assistance in impedance measurements.\\

\end{document}